\documentclass[11pt,a4paper]{article}

\usepackage{enumerate,epsfig,psfrag,graphicx}
\usepackage{amsfonts,amsmath,amssymb}
\usepackage{MnSymbol,arydshln}

\DeclareMathAlphabet{\mathpzc}{OT1}{pzc}{m}{it}

\newcommand{\DIV}[1][\normalsize]{\,\mbox{div}\,}

\newcommand{\wt}[1]{\widetilde{#1} }

\newcommand{\raiseformprod}[2]{\raisebox{-0.4\height}{$#1\invbackneg$}}
\newcommand{\formprod}{\mathpalette\raiseformprod\relax}
\newcommand{\LieDer}[2]{{\mathfrak{L}_{#1} #2}}

\def\N{ {\mathbb N} }
\def\R{ {\mathbb R} }

\def\dd{ \textit{\textrm{d}} }

\def\beq{\begin{equation}}
\def\eeq{\end{equation}}
\def\bea{\begin{eqnarray}}
\def\eea{\end{eqnarray}}

\begin{document}
\renewcommand{\thefootnote}{$\blacklozenge$}

\overfullrule=0pt
\parskip=2pt
\parindent=12pt
\headheight=0in \headsep=0in \topmargin=0in \oddsidemargin=0in

\vspace{ -3cm} \thispagestyle{empty} \vspace{-1cm}
\begin{flushright} 
\footnotesize
\tt{ITP-UU-13/27}
\end{flushright}%

\begin{center}
\vspace{1.2cm}
{\Large\bf \mathversion{bold}
Logarithmic two-Point Correlation Functions from a $z = 2$ Lifshitz Model
}

\vspace{0.8cm} {
 
T.~Zingg\footnote{T.Zingg@uu.nl}
}\\
\vskip  0.5cm

\small
{\em

Institute for Theoretical Physics and Spinoza Institute\\Universiteit Utrecht,  3584 CE Utrecht, The Netherlands}
\normalsize
\end{center}

\begin{abstract}
The Einstein--Proca action is known to have asymptotically locally Lifshitz
spacetimes as classical solutions.
For dynamical exponent $z=2$,
two-point correlation functions for fluctuations
around such a geometry are derived analytically. 
It is found that the retarded correlators are stable in the sense that all quasinormal
modes are situated in the lower half-plane of complex frequencies.
Correlators in the longitudinal channel exhibit features
that are reminiscent of a structure usually obtained in field theories
that are logarithmic, i.e.~contain an indecomposable highest weight representation.
This suggests the model at hand
as a candidate for a gravity dual
of a logarithmic field theory with anisotropic scaling symmetry.
\end{abstract}

\newpage
\thispagestyle{empty}
\tableofcontents
\newpage

\renewcommand{\thefootnote}{\arabic{footnote}}
\setcounter{footnote}{0}
\setcounter{page}{1}

\section{Introduction}

In condensed matter physics, Lifshitz -- or anisotropic --
scaling symmetry is expected to arise
in several phenomena that involve phase transition, 
in particular close to quantum critical points~\cite{PhysRevB.14.1165,2005Natur.433..226C}.
Though several features are well understood,
prevalent methods in field theory often proved unsatisfying when
trying to investigate such systems that are governed by strong interaction.
Due to its weak/strong duality, the AdS/CFT correspondence~\cite{Maldacena:1997re}
provides an intriguing way to 
approach these problems from a new angle and
has developed during the last decade into a rich field of research --
see~\cite{Hartnoll:2009sz,McGreevy:2009xe,Sachdev:2010ch} for reviews.
An idea to use this framework for problems with anisotropic scaling
was presented in~\cite{Kachru:2008yh}, where starting from a metric 
that incorporates Lifshitz invariance
a gravitational dual for the description of critical phenomena was conjectured.
Since then, several gravity models that contain spacetimes with such a
symmetry -- at least asymptotically or locally -- were analyzed.
In particular, models with a dynamical exponent $z = 2$,
which will be in the focus of this paper,
are in the mean time quite well understood.
These models were originally derived from 
a more phenomenological and bottom-up point of view,
but recent years have seem many
successful ways to realize them as well by a top-down approach via an
embedding into a string theory framework,
see e.g.~\cite{Balasubramanian:2010uk,Donos:2010tu,Cassani:2011sv,Chemissany:2011mb,Chemissany:2012du}
to just name a few.
This provides stronger evidence that there is a consistent and
well-defined way to establish a gauge/gravity correspondence in these systems.

The following deals with $z = 2$ Lifshitz solutions of the equations of motion for an 
Einstein--Proca -- or massive vector -- action~\cite{Taylor:2008tg}.
It was shown in~\cite{Chemissany:2011mb,Chemissany:2012du} that this model
can be uplifted to a string theory framework and, furthermore,
that there is a procedure to holographically renormalize this action
even in the presence of a logarithmic divergence, which will be crucial later on.
Thus, for simplicity, calculations will be performed in the four-dimensional
bulk action.
It will be argued that the aforementioned logarithmically diverging mode can be
identified as the source of a logarithmic partner of the energy,
in a sense similar to how the concept was introduced in
in a CFT context~\cite{Gurarie:1993xq}.
Schematically, these are theories that contain 
representations
that are indecomposable but
not diagonalizable and hence contain non-trivial Jordan cells.
This has several consequences for the properties of the fields involved,
in particular, a tower of
logarithmic terms in correlation functions appears according to a specific scheme.\footnote{
It is this feature that in this paper will be used as a criterion 
to call a field theory logarithmic.}
Since they were put on a solid theoretical footing,
LCFTs have found a plethora of applications, such as
gravitationall dressed CFTs, (multi)critical polymers, percolation, 
two-dimensional (magnetohydrodynamic) turbulence,
the (fractional) quantum Hall effect, the sandpile model
and disordered systems.
A comprehensive overview of the theory of LCFTs as well as detailed references to
the aforementioned application can be found 
in~\cite{Flohr:2001zs,Gaberdiel:2001tr,Creutzig:2013hma}.
In the absence of conformal symmetry, the study of field theories 
with such logarithmic features has not yet enjoyed much attention.
First attempts to extend these features to models with Lifshitz scaling
symmetry were made~\cite{Bergshoeff:2011xy}, but in order to obtain
an understanding on the same level as LCFT's there is still a lot
more work to be done.

The paper is organized as follows.
Sec.~\ref{sec:setup} contains a short review of
the notation of an asymptotically Lifshitz fixed point of 
Einstein Gravity coupled to a Proca Field and a summary of results and features 
that were obtained so far.
In sec.~\ref{sec:lin} follows an analysis of the problem in a linearized approximation.
This is sufficient to extract the necessary information to
calculate two-point functions for the different modes involved.
These modes can be split into transversal and longitudinal modes,
often referred to as the shear and sound channel.
A treatise of the shear modes follows in sec.~\ref{sec:shear},
the one for sound modes in sec.~\ref{sec:sound}.
In both cases, quasi-normal frequencies will have a negative imaginary part,
indicating that the system is stable. 
In addition, the two-point correlation functions in the sound channel
are found to exhibit features of a logarithmic field theory.

\section{Einstein Gravity with Proca Field}
\label{sec:setup}

As mentioned in the introduction, the following treatise deals with a bulk action consisting of Einstein gravity coupled minimally to a Proca field,
\bea
\mathpzc{S}
	&=&	\mathpzc{S}_{EH} + \mathpzc{S}_{Proca}	\; ,	 \label{eq:bulk_action}	\\
\mathpzc{S}_{EH}
	&=&	\frac{1}{2 \kappa} \int (R-2\Lambda) \mathit{v}		\; ,	\\
\mathpzc{S}_{Proca}
	&=&	-\frac{1}{4 \kappa} \int \left( \dd P \wedge * \dd P + \mathit{c}\,P\wedge *P\right)		\; .
\eea
The variation of this action leads to the equations of motion,
\bea
G_{\mu\nu} + \Lambda g_{\mu\nu}	&=&	T^P_{\mu\nu}	\; ,	\label{eq:Einstein}	\\
d*dP	&=&	-\mathit{c} *P	\; ,	\label{eq:eom_P}
\eea
with the stress tensor for the Proca field,
\bea
T^P_{\mu\nu}	&=&	\frac{1}{2}\left(	P_{\mu}P_{\nu}
						+ [dP]_{\mu\lambda}{[dP]_{\nu}}^{\lambda}
						-\frac{1}{2}P_{\lambda}P^{\lambda} g_{\mu\nu}
						-\frac{1}{4}[dP]_{\lambda\kappa}[dP]^{\lambda\kappa} g_{\mu\nu}
			\right)\; .	\label{eq:T_P}
\eea
As has become standard when investigating asymptotically Lifshitz
fixed points of this theory, $c$ and $\Lambda$ can be parameterized as,\footnote{An overall length scale has been omitted for simplicity.}
\beq
c		=	2 z		\; ,	\quad 
\Lambda	=	- \frac{z^2+z+4}{2}		\; .
\eeq
Apart from asymptocically AdS spacetimes,
(\ref{eq:Einstein},\ref{eq:eom_P}) is also solved by the so-called
Lifshitz spacetime~\cite{Kachru:2008yh,Taylor:2008tg} with dynamical
exponent $z$, where a tetrad and the massive vector field can be parameterized as follows,
\beq
e^0	=	\frac{1}{r^z}\, \dd t	\; ,	\quad
e^1	=	\frac{1}{r}\, \dd x_1	\; ,	\quad
e_2	=	\frac{1}{r}\, \dd x^2	\; ,	\quad
e^3	=	-\frac{1}{r}\, \dd r		\; ,	\quad
P	=	\sqrt{\frac{2(z-1)}{z}} \frac{1}{r^z} \,\dd t	\; .
\label{eq:Lifshitz_FP}
\eeq
It is called Lifshitz because the metric associated with it,
\bea
\dd s^2	&=&
	- \frac{\dd t^2}{r^{2z}}\, 
		+ \frac{ \dd x_1^2 + \dd x_2^2 +  \dd r^2 }{r^2}\, 
	\; ,
\eea
is left invariant under so-called Lifshitz rescaling,
\bea
t \to \lambda^z t		\; ,	\quad
x_i \to \lambda^z x_i	\; ,	\quad
r \to \lambda^z r		\; .	\label{eq:Lifshitz_scaling}
\eea
Metrics that asymptotically, i.e.~for $r \to 0$, approach the structure
of \eqref{eq:Lifshitz_FP} can be called asymptotically Lifshitz fixed points
of (\ref{eq:Einstein},\ref{eq:eom_P}) and were already studied 
thoroughly in previous work.
What follows next is mainly a summary of known results,
the reader familiar with this work can likely skip this and go ahead to the next section.

\subsection{Asymptotically Lifshitz}
\label{sec:asylif}

In more general terms, an asymptotically locally Lifshitz spacetime could be characterized as follows~\cite{Ross:2009ar,Ross:2011gu}.
The main conditions are already quite explicitly suggested in \eqref{eq:Lifshitz_FP} and
\eqref{eq:Lifshitz_scaling}, i.e.~that the timelike component of the tetrad
must have a different asymptotic scaling than the spacelike ones.
Thus, a metric can be called asymptotically (locally) Lifshitz with dynamical exponent $z$
if there is a function $r$
with spatial infinity at $r=0$ and a semi-orthonormal basis
$e^A$ such that,
\beq
\lim\limits_{r\to 0} \, r^z e^0 = \bar{e}^0 	\; , \qquad
\lim\limits_{r\to 0} \,r\, e^j = \bar{e}^j		\; ,
\label{eq:Lifshitz_cond}
\eeq
with well defined $\bar{e}^j$.
Additional, more subtle conditions can arise
when more stringent conditions on the boundary geometry are imposed.
In particular, demanding that there must be a global identification of the time direction
imposes the condition that $e^0$ must give rise to a foliation of the boundary manifold
into surfaces that are orthogonal to a globally defined timelike vector field.
In addition to \eqref{eq:Lifshitz_cond},
this would also imply that 
there is a global time-coordinate $t$ and a well-defined function
$\bar{\Omega}$, such that,
\beq
\bar{e}^0 = \bar{\Omega} \dd t	\; .
\label{eq:Lifshitz_cond_t}
\eeq
A violation of this condition means that there is not any more 
a clear splitting of the boundary manifold
into a timelike and a spacelike part.
The solution \eqref{eq:Lifshitz_FP} clearly satisfies \eqref{eq:Lifshitz_cond_t},
but a general fluctuation around that fixed point might very well
cause instabilities in the sense that they drive the system
towards a different class of boundary geometries.
When just considering infinitesimal perturbations in order to calculate
correlation functions these issues might seem less relevant,
it should be kept in mind though, that the degrees of freedom --
and thus the number of operators in the dual field theory --
might get reduced when only variations are considered that
stay on a class of solutions that satisfy \eqref{eq:Lifshitz_cond_t}.

\begin{figure}[ht]
\begin{center}
\includegraphics[height=.4\paperwidth]{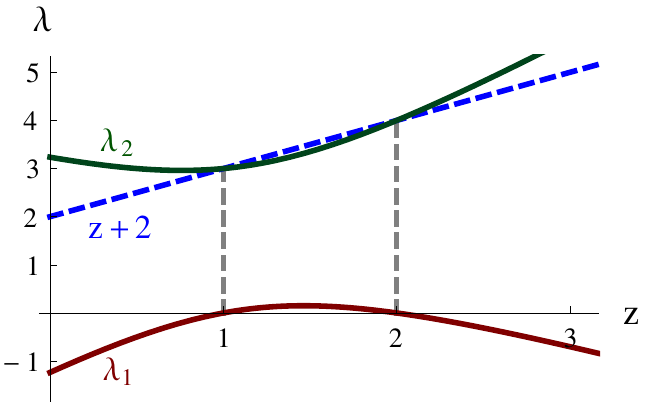}
\caption{\label{fig:Lifshitz_modes} The exponents $\lambda_{1}$ and $\lambda_{2}$ as 
functions of the dynamical exponent $z$.
$\lambda_{1}$ is only positive for $1\leq z \leq 2$, indicating that the
corresponding mode becomes non-renormalizable outside this range.}
\end{center}
\end{figure}

The most straightforward way to study the behavior of such an
asymptotically Lifshitz solution is to
make a general ansatz,
\bea
e^0	=	\frac{\bar{\Omega}}{r^z}\, \dd t
			+ \frac{\bar{\Sigma}}{r^z}		\; ,	\quad
e^j	=	\frac{\bar{\Theta}^j \dd t}{r} + \frac{\bar{e}^j}{r}		\; ,	\quad
e^3	=	\frac{\dd r}{r}		\; ,	\quad
P	=	\frac{\bar{a}}{r^z} \, \dd t
			+ \frac{\bar{b}}{r^z}
			+ \frac{\bar{c}\,\dd r}{r^{z-1}}	\; , \label{eq:general_ansatz}
\eea
with functions $\bar{\Omega}, \bar{a}, \bar{c}$ and
spacelike $2$-forms $\bar{\Sigma}, \bar{\Theta}^j, \bar{e}^j, \bar{b}$,
then plug it into the equations of motion (\ref{eq:Einstein},\ref{eq:eom_P})
and solve recursively order by order as a genrealized power series in $r$,
analogous to the Fefferman--Graham expansion of asymptotically
AdS spacetimes~\cite{FeffermanGraham}.
Schematically, using $\xi$ to represent a generic degree of freedom,
this series will be of the form,\footnote{
For illustrative purposes in the following discussion and
the case $z\to 2$, the sum in \eqref{eq:xi_expansion_general} is expressed
explicitly with $\lambda_{1}$ and $\lambda_{2}$, but it should
be noted that this formally introduces a certain redundancy since
$\lambda_{1}+\lambda_{2} = z+2$.}
\beq
\xi 	=	r^\Delta \sum\limits_{k,l,m,n \geq 0} \xi^{(k,l,m,n)} r^{2 k+2 l z+m \lambda_1+n \lambda_2}	\; ,
\label{eq:xi_expansion_general}
\eeq
where $\Delta$ denotes a certain exponent that depends on which
mode was picked and,
\bea
\lambda_{1,2} &=&	\frac{1}{2}\left( z+2 \mp \sqrt{9z^2 - 20z + 20}  \right)\; .
\eea
A plot of how $\lambda_{1,2}$ behave as a function of $z$ is given
in fig.~\ref{fig:Lifshitz_modes}.
For $1 < z < 2$ all exponents on the right hand side
of \eqref{eq:xi_expansion_general} are positive,
when $z > 2$ the exponent $\lambda_1$ is negative and the associated
mode needs to be tuned to zero because it would otherwise
drive the system away from the asymptotically Lifshitz fixed point.
A quite comprehensive analysis of the degrees of freedom can be found
in~\cite{Ross:2009ar,Ross:2011gu}.
Summarized,
\begin{center}
\begin{tabular}[ht]{lc p{.66\textwidth}}
$\bullet \quad \Delta = 0$			&: \hspace{5mm}&
	components of a triad describing the geometry in the boundary field theory,
		these are interpreted as sources for
		an energy density $\mathpzc{E}$,
		energy flux $\epsilon^j$
		and stress tensor $\Pi_a^j$	\\
$\bullet \quad \Delta = z+2$		&: \hspace{5mm}&
	expectation values for $\mathpzc{E}$, $\epsilon^j$ and $\Pi_a^j$	\\
$\bullet \quad \Delta = 2(z-1)$		&: \hspace{5mm}&
	two degrees of freedom, which are interpreted as the source for a
		vector operator in the boundary field theory --
		the latter was identified with
		the momentum $\mathpzc{p}_a$ in~\cite{Ross:2009ar}	\\
$\bullet \quad \Delta = 3z$			&: \hspace{5mm}&
	expectation value for $\mathpzc{p}_a$	\\
$\bullet \quad \Delta = \lambda_1$	&: \hspace{5mm}&
	a single degree of freedom, interpreted as source of a scalar operator $\mathpzc{O}_P$
	in the boundary field theory -- 
	or the expectation value of an operator $\wt{\mathpzc{O}}_P$
	in alternative quantization
		\\
$\bullet \quad \Delta = \lambda_2$	&: \hspace{5mm}&
	expectation value for $\mathpzc{O}_P$ --
	or source for $\wt{\mathpzc{O}}_P$
	in alternative quantization	\\	
\end{tabular}
\end{center}
What can be noted is that $\bar{b}$ does not source a vector operator in the 
boundary by itself, but it combines with $\bar{\Sigma}$ to provide sources
for independent $\epsilon^j$ and $\mathpzc{p}_a$.
The same can be said about $\bar{a}$ and $\bar{\Omega}$,
which combine to source $\mathpzc{E}$ and $\mathpzc{O}_P$,
respectively $\wt{\mathpzc{O}}_P$.

Once the degrees of freedom are identified, the question is
which fluctuations are stable and which ones would drive the system away 
from the Lifshitz fixed point.
As already mentioned above, there is obviously an instability
if $\lambda_1 < 0$ since the backreaction
to the geometry would fail to satisfy \eqref{eq:Lifshitz_cond}
unless the $\lambda_1$-mode is fine-tuned to zero.
On the field theory side, this condition means that 
only alternative quantization is viable and
$\langle \wt{\mathpzc{O}}_P \rangle = 0$.
A further reduction of the allowed degrees of freedom might come
from considering Green's functions of fluctuations
around a Lifshitz background and demanding that all quasinormal frequencies
must have negative imaginary part.
For $z>2$ such instabilities were found in certain cases~\cite{Andrade:2013wsa},
but explicit calculation for $z=2$ in the following will reveal
that all quasinormal frequencies are in the lower half-plane.

\subsection{Stress - Energy Complex}
\label{sec:stress_energy}

According to the standard dictionary in gauge/gravity duality, 
a stress-energy tensor for the dual field theory is sourced by the boundary metric.
Given that the dual in the case at hand is expected
to have Lifshitz scaling symmetry
and to be non-relativistic,
it seems less convenient to work with a tensor. 
Instead, a formalism is adopted that uses a tetrad instead of a
metric~\cite{Ross:2009ar,Zingg:2011cw}.
First, use $\star$ to denote the Hodge star operator on each of the slices
with constant $r$ and consider,
\beq
\star\pi_P = \frac{\delta \mathpzc{S}}{\delta  P}				\; ,	\quad
\star\tau^A = \eta_{AB}\frac{\delta \mathpzc{S}}{\delta  e^B}	\; .
\label{eq:tau_def}
\eeq
The former is simply the dual momentum of $P$, the latter
can be seen as the dual momenta of the tetrad $e^A$.
In~\cite{Ross:2009ar},
the frame $e^A$ was chosen such that $P = \xi e^0$ for some function $\xi$.
That choice conveniently associates the components of $e^A$
with the sources of a stress-energy-complex and reduces the remaining degrees of freedom
from the vector $P$ to those of a scalar field,
as was already indicated in the previous section.
This allows to define a stress-energy complex consisting of
energy, energy flux, momentum and stress,
\beq
\mathpzc{E}		\longleftrightarrow	\tau^0(\partial_t)				\; ,	\quad
\epsilon^j	\longleftrightarrow	\tau^0\Bigr|_{\partial_t^\perp}		\; ,	\quad
\mathpzc{p}_a	\longleftrightarrow	\tau^a(\partial_t)				\; ,	\quad
\Pi_a^j			\longleftrightarrow	\tau^a\Bigr|_{\partial_t^\perp}	\; .	\label{eq:stress_energy_complex}
\eeq
These quantities are not completely independent, but subject to some constraints,
\bea
\langle \dot{\mathpzc{E}} + \mathrm{div}\,\epsilon \rangle
	&=&	0	\; ,	\label{eq:energy_conservation}		\\
\langle \dot{\mathpzc{p}} + \mathrm{div}\,\Pi \rangle
	&=&	0	\; ,	\label{eq:momentum_conservation}	\\
\langle z \mathpzc{E} + \mathrm{tr}\,\Pi \rangle
	&=&	0	\; .	\label{eq:trace_cond}				
\eea
The first two are conservation equations,
the last one can be considered as the Lifshitz-equivalent of 
what in CFT would be having a vanishing trace of the stress-energy tensor.
The correspondence \eqref{eq:stress_energy_complex} may at first sight suggest
that it is necessary to choose a specific frame to define a dictionary.
This is not really the case.
While \eqref{eq:stress_energy_complex} gives a clear physical interpretation of the
actual degrees of freedom of the system, it would be possible to proceed,
without major circumstances, to define a dictionary for a generic frame --
which seems sensible, given that the latter is only determined by the metric
up to an $O(1,3)$ transformation.
What however must be kept in mind is a condition that was already implicitly used
in the frame in~\cite{Ross:2009ar}.
Namely, in order to ensure to stay on the space of asymptotically 
Lifshitz solutions it is not allowed to vary all components of $e^A$ and $P$
independently.
In other words, when choosing a generic parameterization, for example \eqref{eq:general_ansatz},
then requiring that the ansatz satisfies (\ref{eq:Einstein},\ref{eq:eom_P}) and
\eqref{eq:Lifshitz_cond} simultaneously 
leads to on-shell conditions that also relate the values of the fields at spatial infinity,
\beq
\mathfrak{X}\left[\bar{\Omega}, \bar{\Sigma}, \bar{\Theta}, \bar{e}^i, \bar{a}, \bar{b},\bar{c}\right]\Bigr|_{r=0}
	= 0	\; .
\label{eq:Xcond}
\eeq
Once the degrees of freedom have been identified,
the standard gauge/gravity dictionary
can be used when considering derivatives with respect to the degrees of freedom
instead of the components of $e^A$ and $P$ directly.
E.g.~with the example \eqref{eq:general_ansatz}, 
$\bar{\Omega}$ can be identified as the source for the energy $\mathpzc{E}$,
which can be seen through boosting and rotating the frame $e^A$ such that $e^0 \parallel P$.
Then, it is straightforward to calculate the expectation value either in a direct
way or by doing a variation under condition \eqref{eq:Xcond} and using Leibnitz's rule,
\beq
\langle \mathpzc{E} \rangle
	= \star \frac{\delta \mathpzc{S}}{\delta \bar{\Omega}} 
	= \star \left[
				\frac{\partial e^0}{\partial\bar{\Omega}} \wedge \star\tau^0
				+\frac{\partial e^1}{\partial\bar{\Omega}} \wedge \star\tau^1 
				+\frac{\partial e^2}{\partial\bar{\Omega}} \wedge \star\tau^2
				+\frac{\partial P}{\partial\bar{\Omega}} \wedge \star\pi_P 
			\right]\Biggr|_{\mathfrak{X}}	\; .
\eeq

\subsection{The Case z = 2 and Logarithmic Field Theories}
\label{sec:z=2}

As already could be seen in fig. \ref{fig:Lifshitz_modes},
the case $z = 2$ is of particular interest and was 
thus also subject to some more detailed investigations~\cite{Chemissany:2012du,Baggio:2011cp,Griffin:2011xs,Baggio:2011ha}.
At this value, $\lambda_1 = 0$ and the mode associated with it becomes marginal.
Furthermore, $\lambda_2 = 4$ and thus the operator $\mathpzc{O}_P$ has the same scaling
dimension as $\mathpzc{E}$ and $\Pi$.
On the gravity side, the general Fefferman--Graham expansion \eqref{eq:xi_expansion_general}
is modified to contain logarithmic terms,
\beq
\xi 	=	r^\Delta \sum\limits_{n \geq 0} \xi^{(n)} r^{2 n}
				+ r^\Delta \log r \,\sum\limits_{n \geq 0} \wt{\xi}^{(n)} r^{2 n}
					+ \ldots	\; ,
\label{eq:xi_expansion_general_z=2}
\eeq
where the dots stand for higher powers in logarithms.
The degrees of freedom are the same as for $z \neq 2$,
only now the ones associated to $\lambda_{1,2}$
have become degrees of freedom in the logarithmic terms
at order $0$ and $4$ respectively.
Furthermore, having $z+2$ being a multiple of $2$ also results in anomalous terms
in the expectation values for the stress-energy complex,
i.e.~(\ref{eq:energy_conservation}-\ref{eq:trace_cond})
can contain non-zero terms on the right hand side,
\bea
\langle\dot{\mathpzc{E}} + \mathrm{div}\,\epsilon \rangle
	&=&	\mathpzc{A}_1	\; ,	\label{eq:energy_conservation_z=2}		\\
\langle\dot{\mathpzc{p}} + \mathrm{div}\,\Pi \rangle
	&=&	\mathpzc{A}_2	\; ,	\label{eq:momentum_conservation_z=2}	\\
\langle 2 \mathpzc{E} + \mathrm{tr}\,\Pi \rangle
	&=&	\mathpzc{A}_3	\; .	\label{eq:trace_cond_z=2}				
\eea
A analysis of the anomalous terms $\mathpzc{A}_j$ is a somewhat tedious affair
and for further details will therefore be referred 
to~\cite{Chemissany:2012du,Baggio:2011cp,Griffin:2011xs,Baggio:2011ha}.
The remainder of the paper deals with two-point correlation functions
at the Lifshitz fixed point
in the presence of the aforementioned logarithmic terms.
The leading order term was often discarded in an analysis
of $z=2$ Lifshitz spacetimes so far,
as it fails to satisfy \eqref{eq:Lifshitz_cond}.
Being logarithmic, it is however a rather mild violation of the
asymptotically Lifshitz condition.
Such logarithmic deformations have already enjoyed a range of studies in
asymptotically AdS spacetimes, like
topologically massive gravity (TMG)~\cite{Skenderis:2009nt,Grumiller:2009mw}
new massive gravity (NMG)~\cite{Grumiller:2009sn} and
other higher-derivative gravity theories~\cite{Johansson:2012fs,Bergshoeff:2012ev}.
They were argued to be duals of a logarithmic conformal field theory
(LCFT)~\cite{Gurarie:1993xq},
which is distinguished from ordinary CFT in that it contains
an indecomposable but not diagonalizable highest weight representation.\footnote{for a recent review of the AdS/LCFT correspondence
see e.g.~\cite{Grumiller:2013at}}
This means that
some of the highest weight states form a non-trivial Jordan cell,
the size of which is usually called the rank of the
indecomposable representation.
A comprehensive review of LCFT's  and their features can for example
be found in~\cite{Flohr:2001zs,Gaberdiel:2001tr,Creutzig:2013hma}.
What will be relevant in the following is that the presence of a
Jordan cell will cause the appearance of logarithmic terms in 
correlation functions.
For example, in rank two, assume $\Psi$ is a field of weight $h$ and 
$\wt{\Psi}$ is its (first) logarithmic partner.
Then, under some general assumptions about the form of the OPE,
two-point correlation functions look as follows,
\bea
\langle \Psi(z) \Psi(0) \rangle				&=&	0	\; ,	\label{eq:LCFT_2pt_1}	\\
\langle \wt{\Psi}(z) \Psi(0) \rangle		&=&	\frac{a}{z^{2h}}	\label{eq:LCFT_2pt_2}		\; ,	\\
\langle \wt{\Psi}(z) \wt{\Psi}(0) \rangle	&=&	\frac{-2a \log z + b}{z^{2h}}	\label{eq:LCFT_2pt_3}	\; ,
\eea
for some constants $a,b$ that depend on the normalization of the fields.
In the absence of conformal symmetry,
the study of field theories 
with such logarithmic features has not yet enjoyed much attention.
First attempts to extend these features to models with Lifshitz scaling
symmetry were made~\cite{Bergshoeff:2011xy}, but a thorough
understanding of these kind of systems is still in its infancy.
Nevertheless, a number qualitative results are at hand
and by comparing those to the two-point correlation functions
that will be calculated
in section sec.~\ref{sec:log}, evidence will be
unveiled that suggests the massive vector model \eqref{eq:bulk_action}
at the $z = 2$ Lifshitz fixed point as a candidate
for a gravity dual to a field theory where the energy operator 
$\mathpzc{E}$ and a scalar operator $\mathpzc{O}_P$
combine to form a logarithmic pair.

\section{Linearization}
\label{sec:lin}

A general solution to the system (\ref{eq:Einstein},\ref{eq:eom_P}) is
not known, but according to the prescription in~\cite{Son:2002sd},
it is sufficient to linearize the equations of motion around
\eqref{eq:Lifshitz_FP} in order to calculate two-point correlation functions
in that background.
A perturbation around that fixed point can be parameterized as
\bea
\delta e^0	&=&	\frac{\Omega}{2 r^2}\, \dd t
			+ \frac{\Sigma_1}{2 r^2}\, \dd x_1
			+ \frac{\Sigma_2}{2 r^2}\, \dd x^2	\;,	\label{eq:pert_param_first}	\\
\delta e^1	&=&	\frac{h_{11}}{2r}\, \dd x_1
			+ \frac{h_{12}}{2r}\, \dd x^2	\;,	\\
\delta e^2	&=&	\frac{h_{12}}{2r}\, \dd x_1
			+ \frac{h_{22}}{2r}\, \dd x^2	\;,	\\
\delta e^3	&=&	0	\;,	\\
\delta P	&=&	\frac{a}{r^2} \, \dd t
			+ \frac{b_1}{r^2}\, \dd x_1
			+ \frac{b_2}{r^2}\, \dd x^2
			+ \frac{c}{r} \, \dd r	\; . \label{eq:pert_param_last}
\eea
What can also be seen, with regard to the general parameterization \eqref{eq:general_ansatz},
is that $\bar{\Theta}^j = 0$, i.e.~there are no timelike components in $e^j$.
Since $e^A$ is only defined up to a $O(1,3)$ transformation,
this choice is possible without loss of generality.

\subsection{Linearized Equations}
\label{sec:lineq}

Plugging the ansatz (\ref{eq:pert_param_first}-\ref{eq:pert_param_last}) into the equations of motion (\ref{eq:Einstein},\ref{eq:eom_P})
and going to Fourier space leads to a set of linear ODEs for the perturbations,
\bea
  r^2 \Omega''- 4 r \Omega' +\left(-\frac{k^2 r^2}{2} +6\right) \Omega 
  -\frac{r}{2} h_{11}'
     + \left(\frac{\omega^2 r^4}{2}+\frac{k_{2}^2 r^2}{2}\right) h_{11}
      \hspace{14mm} & & \nonumber	\\
  -\frac{r}{2} h_{22}' + \left(\frac{\omega^2 r^4}{2}+\frac{k_{1}^2 r^2}{2}\right)h_{22}
  -k_{1} k_{2} r^2 h_{12}
  + 3 r a'-12 a \hspace{1cm} & & \nonumber	\\
  -\frac{\omega k_{1} r^2}{2} \Sigma_{1}
  -\frac{\omega k_{2} r^2}{2} \Sigma_{2}
  +3 i r^2 \omega c
  & =& 0	\; , \label{eq:lineom_1}	\\
  r^2 a''-3 r a'+r^2  \left(-k_{1}^2-k_{2}^2\right)a
  +r \Omega'
  -r h_{11}'
  -r h_{22}'  \hspace{32mm} & & \nonumber	\\
  -k_{1} r^2 \omega b_{1}
  -k_{2} r^2 \omega b_{2}
  +i r^2 \omega c'
    & =& 0	\; , \label{eq:lineom_2}	\\
  r^2 h_{11}''-\frac{5}{2} r h_{11}'+\left(\frac{\omega^2 r^4}{2}-\frac{k_{2}^2 r^2}{2}\right) h_{11}
  +\left(\frac{\left(k_{2}^2-k_{1}^2\right)r^2}{2} - 2\right) \Omega	 \hspace{14mm} & & \nonumber	\\
  -r a'+4 a	\nonumber
  +\frac{r}{2} h_{22}'+\left(-\frac{\omega^2 r^4}{2}-\frac{k_{1}^2 r^2}{2}\right) h_{22}
  +k_{1} k_{2} r^2 h_{12}	\hspace{10mm} & & \nonumber	\\
  -\frac{\omega k_{1} r^2}{2} \Sigma_{1}
  +\frac{\omega k_{2} r^2}{2} \Sigma_{2}
  -i r^2 \omega c
    & =& 0	\; , \label{eq:lineom_3}	\\
  r^2 h_{22}''-\frac{5 r}{2} h_{22}'+ \left(\frac{\omega^2 r^4}{2}-\frac{k_{1}^2 r^2}{2}\right)h_{22}
  +\left(\frac{\left(k_{1}^2-k_{2}^2\right)r^2}{2}-2\right)\Omega \hspace{14mm} & & \nonumber	\\
  -r a'+4 a
  +\frac{r}{2} h_{11}'+ \left(-\frac{\omega^2 r^4}{2}-\frac{k_{2}^2 r^2}{2}\right) h_{11}
  +k_{1} k_{2} r^2 h_{12}	\hspace{10mm} & & \nonumber	\\
  +\frac{\omega k_{1} r^2}{2} \Sigma_{1}
  -\frac{\omega k_{2} r^2}{2} \Sigma_{2}
  -i r^2 \omega c
    & =& 0	\; , \label{eq:lineom_4}
\eea
\bea
  r^2 h_{12}''-3 r h_{12}'+\omega^2 r^4 h_{12}
  -k_{1} k_{2} r^2 \Omega
  -\frac{1}{2} k_{2} r^2 \omega \Sigma_{1}
  -\frac{1}{2} k_{1} r^2 \omega \Sigma_{2}
    & =& 0	\; , \label{eq:lineom_5}	\\
  r^2 \Sigma_{1}''-5r \Sigma_{1}' + \left(- k_{2}^2 r^2 + 8 \right) \Sigma_{1}
  -2 k_{1} r^4 \omega h_{22}
  +2 k_{2} r^4 \omega h_{12}
  +k_{1} k_{2} r^2 \Sigma_{2} \hspace{16mm} & & \nonumber	\\
  +4 r b_{1}'- 16 b_{1}
  -4 i k_{1} r^2 c
    & =& 0	\; , \label{eq:lineom_6}	\\
  r^2 \Sigma_{2}''-5r \Sigma_{2}' + \left( - k_{1}^2 r^2 + 8 \right)  \Sigma_{2}
  -2 k_{2} r^4 \omega h_{11}
  +2 k_{1} r^4 \omega h_{12}
  +k_{1} k_{2} r^2 \Sigma_{1} \hspace{16mm} & & \nonumber	\\
  +4 r b_{2}'- 16 b_{2}
  -4 i k_{2} r^2 c
    & =& 0	\; , \label{eq:lineom_7}	\\
  r^2 b_{1}''-5r b_{1}'+ \left( \omega^2 r^4-k_{2}^2 r^2+4 \right) b_{1}
  +\omega k_{1} r^4 a
  +k_{1} k_{2} r^2 b_{2}
  +r \Sigma_{1}' - 2 \Sigma_{1}
  -i k_{1} r^3 c'
  & =& 0	\; , \label{eq:lineom_8}	\\
  r^2 b_{2}''-5r b_{2}'+ \left( \omega^2 r^4-k_{1}^2 r^2+4\right) b_{2}
  +\omega k_{2} r^4 a
  +k_{1} k_{2} r^2 b_{1}
  +r \Sigma_{2}' - 2 \Sigma_{2}
  -i k_{2} r^3 c'
  & =& 0	\; , \label{eq:lineom_9}	\\
  \left(-\omega^2 r^4 + k^2 r^2 + 4 \right) c
  +i \omega r^2 \Omega
  +i \omega r^3 a'- 2i \omega r^2 \omega a
  -i \omega r^2 h_{11}
  -i \omega r^2 \omega h_{22} \hspace{3mm} & & \nonumber	\\
  +i k_{1} \Sigma_{1}
  +i k_{2} \Sigma_{2}
  +i k_{1} r b_{1}' - 2 i k_{1} b_{1}
  +i k_{2} r b_{2}' - 2 i k_{1} b_{2}
    & =& 0	\; ,
\label{eq:lineom_10}
\eea
that are subject to some additional constraints,
\bea
    -r \Omega' - \frac{k^2 r^2}{2} \Omega
  -r a'
  -\frac{3 r}{2} h_{11}'+\left(\frac{\omega^2 r^4}{2}-\frac{k_{2}^2 r^2}{2}\right) h_{11}
  +k_{1} k_{2} r^2 h_{12} \hspace{1.2cm}	& &	 \nonumber	\\
  -\frac{3 r}{2} h_{22}'
  +\left(\frac{\omega^2 r^4}{2}-\frac{k_{1}^2 r^2}{2}\right) h_{22}
  -\frac{ \omega k_{1} r^2 }{2} \Sigma_{1}
  -\frac{ \omega k_{2} r^2 }{2} \Sigma_{2}
  -i r^2 \omega  c
  & = & 0	\; ,	\label{eq:constr_1}	\\
  -\frac{i k_{1} r}{2} \Omega'+\frac{1}{2} i k_{1} \Omega
  -i k_{1} a
  +\frac{i k_{2} r}{2} h_{12}'
  -\frac{i k_{1} r}{2} h_{22}'
  -\frac{i  \omega r}{4} \Sigma_{1}' + \frac{i \omega}{2} \Sigma_{1}'
  -i \omega  b_{1}
  & = & 0	\; , \label{eq:constr_2}	\\
  -\frac{ i k_{2} r}{2} \Omega'+\frac{i k_{2}}{2} \Omega
  -i k_{2} a
  -\frac{i k_{2} r}{2} h_{11}'
  +\frac{i k_{1} r}{2} h_{12}'
  -\frac{i  \omega r}{4} \Sigma_{2}' + \frac{i \omega}{2} \Sigma_{2}'
  -i \omega  b_{2}
  & = & 0	\; , \label{eq:constr_3}	\\
  \frac{i \omega r^3}{2} h_{11}'+\frac{i \omega r^2}{2} h_{11}
  +\frac{i \omega r^3}{2} h_{22}'+\frac{i \omega r^2}{2} h_{22}
  -\frac{i k_{1} r}{4} \Sigma_{1}'
  -\frac{i k_{2} r}{4} \Sigma_{2}'
  -2 c
  & = & 0	\; . \label{eq:constr_4}
\eea
It should be mentioned that $c$ does not contain any 
additional degrees of freedom,
it is completely determined by the other functions.
Nevertheless, introducing $c$ in \eqref{eq:pert_param_last}
is crucial, as it would not be possible to solve the
equations with all constraints in a meaningful way otherwise.
To proceed, it is convenient to split the functions involved into
shear and sound modes.
For this purpose,
\beq
h =	\left[\begin{array}{cc}
			h_{11}	& h_{12}	\\
			h_{12}	& h_{22}
		\end{array}\right]	\; , \quad
\mathpzc{J} =	\left[\begin{array}{cc}
				& -1	\\
			1	& 
		\end{array}\right]	\; , \quad
\Sigma =	\left(\begin{array}{cc}
			\Sigma_1	\\
			\Sigma_2
		\end{array}\right)	\; , \quad
b =	\left(\begin{array}{cc}
			b_1	\\
			b_2
		\end{array}\right)	\; , \quad
k =	\left(\begin{array}{cc}
			k_1	\\
			k_2
		\end{array}\right)	\; . \quad
\label{eq:shorthand}
\eeq
Using the shorthand introduced above,
\beq
\check{h}		=	\frac{1}{k^2} k\cdot h \cdot \mathpzc{J} \cdot k	\; , \;\;	
\check{\Sigma}	=	k\cdot \mathpzc{J} \cdot \Sigma	\; , \;\;
\check{b}		=	k\cdot \mathpzc{J} \cdot b	\; ,
\label{eq:shear_modes}
\eeq
span the modes of the shear channel that will be analyzed in
sec.~\ref{sec:shear}.
The sound modes are spanned by
\beq
\Omega \; , \;\;
a \; , \;\;
c \; , \;\;
\mathrm{tr}\,h	\; , \;\;
\hat{h}			=	\frac{1}{k^2} k\cdot h \cdot k	\; , \;\;	
\hat{\Sigma}	=	k\cdot \Sigma	\; , \;\;
\hat{b}			=	k\cdot b	\; ,
\label{eq:sound_modes}
\eeq
and a solution for these modes will be discussed in sec.~\ref{sec:sound}.

\subsection{Renormalization and On-Shell Action}
\label{sec:renorm}

In the standard gauge/gravity dictionary, the on-shell gravity action
is identified with the generating functional for correlation functions in
the dual field theory,
\bea
\langle O_1(x_1) \cdots O_n(x_n) \rangle
	&=&	(-i)^{n+1} \frac{\delta^n \mathpzc{S}}{\delta J_1(x_1)  \cdots \delta J_n(x_n)}	\; .
\label{eq:genfunc}
\eea
As is often the case, the action functional is not finite
for general solutions to the equations of motion and needs to
be renormalized,
\bea
\mathpzc{S}^{ren} &=&
\mathpzc{S}^{bulk} + \mathpzc{S}^{GH} + \mathpzc{S}^{ct} + \mathpzc{S}^{ct, anom}	\; ,
\label{eq:S_ren}
\eea
where $\mathpzc{S}^{GH}$ denotes the usual Gibbosns-Hawking term needed to make the
variation of the action well-defined, and the remaining terms on the right hand side
are needed to make $\mathpzc{S}^{ren}$ finite on-shell.
The term $\mathpzc{S}^{ct,anom}$ denotes anomalous counterterms,
i.e. terms needed to cancel logarithmic divergences that arise
in the bulk action and are expected to appear for even $z$.
These must not be confused with the mechanism behind the appearance
of logarithmic terms
in the correlation functions, which will be the main topic in sec.~\ref{sec:log}.
The former appear due to the dimension of the energy, $z+2$,
being a multiple of $2$.
Those terms do not introduce new degrees of freedom 
and are, in fact, completely determined by
the curvature and other
derivatives of the boundary fields.
The logarithmic terms that  
arise when the exponents of two or more different modes coincide exactly,
however, do contain actual degrees of freedom.
This is the feature that leads to Jordan cells in the decomposition of
the on-shell solution into Eigenstates with respect to the normal Lie 
derivative $\mathfrak{L}_n$.
The expectation is that this is related to a similar
structure of Jordan cells in representations
of the field theory dual.

A systematic procedure to construct $\mathpzc{S}^{ct}$
and $\mathpzc{S}^{ct, anom}$ is holographic renormalization,
where an on-shell solution is expanded into a Fefferman--Graham expansion
in a radial coordinate $r$ and counterterms are calculated order by order
-- see e.g.~\cite{Bianchi:2001kw,Skenderis:2002wp} to get an overview.
This procedure is in the meanwhile well understood for asymptotically
AdS spacetimes.
First attempts to extend these results to asymptotically Lifshitz
spacetime were presented in~\cite{Taylor:2008tg,Ross:2009ar,Zingg:2011cw},
more sophisticated descriptions followed~\cite{Ross:2011gu,Baggio:2011cp,Mann:2011hg}.
A detailed analysis for the case $z=2$ can be found in~\cite{Griffin:2011xs,Baggio:2011ha},
where also much attention was devoted to calculating the anomalous
terms arising in this case.
A maybe somewhat more convenient and elegant way
for a holographic renormalization of a $z = 2$
Lifshitz spacetime was presented in~\cite{Chemissany:2012du},
via an embedding into an asymptotically AdS spacetime where holographic
renormalization is well understood.
This procedure also included the leading logarithmic divergence 
that was discarded in most of the previously mentioned
publications.

For the purpose of this paper, however, 
using the full formalism for holographic renormalization
and calculating anomalous terms seems a bit
like breaking a butterfly upon a wheel.
As already mentioned before, two-point functions can be calculated via
a linearization of the equations of motion.
This means that in order to obtain all necessary information it is
sufficient to know the action up to quadratic order.
Counterterms that render the action finite at this
level of detail can be constructed in a rather straightforward
way -- an explicit expression can be found in app.~\ref{app:cts}.
There is a certain ambiguity in these counterterms,
as they are only determined up to finite contributions
by the requirement to cancel divergences of the action on-shell.
Knowing these contributions would be crucial in order to calculate
anomalous terms in one-point functions, but for
two-point functions they would only contribute as contact terms
and will thus be of little significance in the following.

Renormalization results in an explicit expression
for the on-shell action at quadratic order.
As mentioned above, an asymptotic expansion for
a general solution of (\ref{eq:lineom_1}-\ref{eq:constr_4}) can perturbatively
be found by making a generalized power series ansatz
like~\eqref{eq:xi_expansion_general_z=2} and
demanding that this series solves the equations of motion.
This will result in various conditions
relating the coefficients, which can be calculated order by order.
To find an explicit expression for the on-shell action at the boundary
at quadratic order
it is sufficient to calculate these coefficients up to order $6$ and then inserting the
series into \eqref{eq:S_ren}.
After a somewhat tedious calculation,
the on-shell action can be expressed as,
\bea
\mathpzc{S}^{on-shell}
	&=&	\frac{1}{2\kappa}\frac{1}{(2 \pi)^3} \int \langle X , \mathpzc{M} X \rangle\,\dd \omega\,\dd^2 k
	+ \mathpzc{O}(x_3,(\omega+k^2)^2)	\; .
\label{eq:S_onshell}
\eea
The last term on the right hand side is there to
remind that the on-shell action is determined up to quadratic
order and derivative corrections that would only contribute as
contact terms in two-point functions and will thus not be considered in the following.
Using the notation~\eqref{eq:shorthand}, expressions 
for $X$ and $\mathpzc{M}$ can be given,
\bea
X	&=&	\left(
				\Omega^{(0)},\,
				\wt{\Omega}^{(0)},\,
				\check{\Sigma}^{(2)},\,
				\hat{\Sigma}^{(2)},\,
				\check{\Sigma}^{(0)},\,
				\hat{\Sigma}^{(0)},\,
				\mathrm{tr}\,h^{(0)},\,
				\hat{h}^{(0)},\,
				\check{h}^{(0)},\,
				\check{\mathpzc{u}},\,
				\hat{\mathpzc{v}},\,
				\check{\mathpzc{v}},\,
				T,\,
				\wt{T}
		\right)	\; ,	\label{eq:x_vector} 	\\
\mathpzc{M}	&=&	\frac{1}{2} \left[ \begin{array}{cccccccccccccc}
		\hspace{4mm}	&	&	&	&	&	&	&	&	&	&	&	&	& 1 \\
 		& \hspace{4mm}	&	&	&	&	&	&	&	&	&	&	& 1 &	\\
	 	&	& \hspace{4mm}	&	&	&	&	&	&	&	&	& 1 &	&	\\
 		&	&	& \hspace{4mm}	&	&	&	&	&	&	& 1 &	&	&	\\
	 	&	&	&	& \hspace{4mm}	&	&	&	&	& 1 &	&	&	&	\\
 		&	&	&	&	& \hspace{4mm}	&	&	&	&	&	&	&	& \frac{\omega }{k^2} \\
	 	&	&	&	&	&	& \hspace{4mm}	&	&	&	& \omega  &	&	& -2 \\
 		&	&	&	&	&	&	& \hspace{4mm}	&	&	& -2 \omega  &	&	& 2 \\
	 	&	&	&	&	&	&	&	& \hspace{4mm}	&	&	& 2 \omega  &	&	\\
 		&	&	&	& 1 &	&	&	&	& \hspace{4mm}	&	&	&	&	\\
	 	&	&	& 1 &	&	& \omega  & -2 \omega  &	&	& \hspace{4mm}	&	&	&	\\
 		&	& 1 &	&	&	&	&	& 2 \omega  &	&	& \hspace{4mm}	&	&	\\
 		& 1 &	&	&	&	&	&	&	&	&	&	& \hspace{4mm}	&	\\
	 1	&	&	&	&	& \frac{\omega }{k^2} & -2 & 2 &	&	&	&	&	& \hspace{4mm}
	\end{array}\right]	\; .	\label{eq:M_matrix}	
\eea
In \eqref{eq:x_vector}, the index $(2j)$ refers to the order of the coefficient in the
generalized power series~\eqref{eq:xi_expansion_general_z=2}, the last five terms are,
\beq
\wt{T}
	=	- \wt{\Omega}^{(4)}	\; ,	\quad
T	
	=	2 \Omega^{(4)}
				- \frac{5}{2} \wt{\Omega}^{(4)}	\; ,	\quad
\hat{\mathpzc{v}}
	=	\frac{1}{2 k^2}\hat{\Sigma}^{(4)}	\quad
\check{\mathpzc{v}}
	=	\frac{1}{2 k^2}\check{\Sigma}^{(4)}	\; \quad
\check{\mathpzc{u}}
	=	\frac{3}{2 k^2}\check{\Sigma}^{(6)}	\; .
\eeq
As explained in sec.~\ref{sec:asylif}, the modes 
$\Omega^{(0)}$, $h^{(0)}$, $\Sigma^{(0)}$ and $\Sigma^{(2)}$
can be associated with the sources for
the energy $\mathpzc{E}$,
stress $\Pi$,
energy flux $\epsilon$
and momentum $\mathpzc{p}$.
In accordance with~(\ref{eq:energy_conservation_z=2}-\ref{eq:trace_cond_z=2})
follows,
\bea
\langle\dot{\mathpzc{E}} + \mathrm{div}\,\epsilon \rangle
	&=&	\mathpzc{O}((\omega+k^2)^2)	\; ,	\label{eq:energy_conservation_z=2_quadr}		\\
\langle\dot{\mathpzc{p}} + \mathrm{div}\,\Pi \rangle
	&=&	\mathpzc{O}((\omega+k^2)^2)	\; ,	\label{eq:momentum_conservation_z=2_quadr}	\\
\langle 2 \mathpzc{E} + \mathrm{tr}\,\Pi \rangle
	&=&	\mathpzc{O}((\omega+k^2)^2)	\; .	\label{eq:trace_cond_z=2_quadr}				
\eea
Again, the term on the right hand side is there to remind
that derivative correction are expected, but they will not be
of importance for the following calculations.
Despite the few shortcomings in the calculation of expectation values,
the quadratic on-shell action
\eqref{eq:S_onshell} contains all necessary information to extract
the data for two-point functions.
These correspond to the
Green's functions of the fields in the linearized equations
(\ref{eq:lineom_1}-\ref{eq:constr_4}).
Examining these in the next sections will also allow to
shed some light on the role
of the scalar operator sourced by $\wt{\Omega}^{(0)}$,
which has the same dimension as $\mathpzc{E}$ and $\Pi$.

\section{Shear Channel}
\label{sec:shear}

The shear modes \eqref{eq:shear_modes} can be combined into two master fields,
\beq
\mathpzc{X} = \check{h} +\frac{1}{2\omega r^2}\left( \check{\Sigma} - 2\check{b} \right)	\; ,	\quad
\mathpzc{Y} = \frac{1}{\omega^2 r^3}\left( \check{\Sigma}' - 2\check{b}' \right)	
					- \frac{2}{\omega^2 r^4}\left( \check{\Sigma}' - 2\check{b}' \right)\; .
\eeq
These solve a set of decoupled second order differential equations,
\bea
v^2 \mathpzc{X}'' - v \mathpzc{X}' + v \left(2 v-\lambda\right) \mathpzc{X} &=& 0	\; ,	\label{eq:master_X} \\
v^2 \mathpzc{Y}'' + v \left(v-\frac{\lambda}{2}\right) \mathpzc{Y} &=& 0	\; ,	\label{eq:master_Y}
\eea
which used the substitution of variables,
\beq
v=\frac{r^2}{2 \omega} \; , \quad \lambda = \frac{k^2}{\omega}	\; .	\label{eq:rTOv}
\eeq
A general analytic solution to (\ref{eq:master_X},\ref{eq:master_Y})
can be written as follows,
\bea
\mathpzc{X}
  &=&	A_\mathpzc{X} e^{i v}\,\Gamma \left(1+\frac{i\lambda}{4}\right) U\left(\frac{i \lambda}{4} ,0,- 2 i v\right)
  		+ B_\mathpzc{X} e^{- i v}\,\Gamma \left(1-\frac{i\lambda}{4}\right) U\left(-\frac{i \lambda}{4} ,0, 2 i v\right)	\; ,	\label{eq:master_X_sol}	\\
\mathpzc{Y}
  &=&	A_\mathpzc{Y} e^{i v}\,v^2\,\Gamma \left(\frac{6 + i \lambda}{4} \right) U\left(\frac{6+i \lambda}{4} ,3,- 2 i v\right)
  		+ B_\mathpzc{Y} e^{- i v}\,v^2\,\Gamma \left(\frac{6 - i \lambda}{4} \right) U\left(\frac{6 - i \lambda}{4} ,1,2 i v\right)	\; . \qquad	\label{eq:master_Y_sol}
\eea
In these expressions, $U$ stands for the Tricomi function which is a solution to
the confluent hypergeometric differential equation.
This function has a branch cut in the third argument along the negative real axis.
Explicit expressions for the shear modes follow immediately,
\bea
\check{h} &=&
	\mathpzc{X}
	+ \frac{1}{2} \int \mathpzc{Y} \; ,	\label{eq:chh_expl}		\\
\check{\Sigma} &=&
	- 2 \lambda \mathpzc{X}
	+ 2 \mathpzc{Y} - 2v \mathpzc{Y}' - 2v\int \mathpzc{Y}		\; , \label{eq:chSigma_expl}		\\
\check{b} &=&
	- \lambda \mathpzc{X}
	+ \mathpzc{Y} - v\mathpzc{Y}'	\; .	\label{eq:chb_expl}
\eea
As the intent is to calculate retarded Green's functions, (\ref{eq:master_X_sol},\ref{eq:master_Y_sol})
will be subject to infalling wave condition in the interior, which means
$A_{\mathpzc{X},\mathpzc{Y}} = 0$.
Demanding \eqref{eq:Lifshitz_cond_t} to be satisfied would furthermore
induce a relation between $B_{\mathpzc{X}}$ and $B_{\mathpzc{Y}}$.
However, in order to calculate correlation functions,
this condition will not be imposed on-shell in the following.
Keeping in mind that there is an integration constant in
(\ref{eq:chh_expl}-\ref{eq:chb_expl}),
this leaves three degrees of freedom.
These could be parameterized by $\check{h}^{(0)}$, $\check{\Sigma}^{(0)}$
and $\check{\Sigma}^{(2)}$,
other modes in the shear channel can then be expressed 
via those coefficients,
\bea
\hspace{-7mm}\check{\mathpzc{v}}	&=&
	\frac{ \omega \left[ \psi_1 + 4\ln\left(-i \omega\right)\right] }{32} \, \check{h}^{(0)}
	+ \frac{\omega \left[ \lambda^2 - 4 + \lambda \psi_2 \right] }{64 \lambda} \, \check{\Sigma}^{(0)}	
	+ \frac{ \psi_1 + 4\ln\left(-i \omega\right) }{64} \, \check{\Sigma}^{(2)}\; ,	\label{eq:v_check}	\\
\hspace{-7mm}\check{\mathpzc{u}}	&=&
	\frac{ \omega^2 \psi_2 }{32} \, \check{h}^{(0)}
		+\frac{ \omega^2 \left\{
					24 \left( \lambda^2 + 2 \right) \psi_2
					-\lambda \left( \lambda^2 + 4 \right) \left[
							3\psi_1
							- 28
							+ 12 \ln\left(-i \omega\right)
						\right]
					\right\}
			}{3072 \lambda} \, \check{\Sigma}^{(0)}
	+ \frac{ \omega \psi_2 }{64} \, \check{\Sigma}^{(2)}
	\; . \;	\label{eq:u_check}
\eea
The $\psi_j$ are substitutions for expressions that involve the Euler--Mascheroni constant
$\gamma$ and the digamma function $\psi$,
\bea
\psi_1	&=&
		2 i \lambda - 6 + 8 \gamma
		- \lambda^2 \psi\left(\frac{i \lambda}{4}\right)
		+ \left(\lambda^2+4\right) \psi\left( \frac{i \lambda + 2}{4} \right)
\; ,	\\
\psi_2	&=&
	\frac{1}{4} \left[						
		2 i \lambda^2 - 4 \lambda^2 + 8 i
		- \lambda \left(\lambda^2+4\right) \psi\left( \frac{i \lambda}{4} \right)
		+ \lambda \left(\lambda^2+4\right) \psi\left( \frac{i \lambda + 2}{4} \right)
	\right]	\; .
\eea
Following the standard prescription~\cite{Son:2002sd}
it is now straightforward to obtain the retarded Green's functions,
respectively two-point correlation functions,
for the operators in the shear channel from these expressions.
These operators are the transversal components of 
the stress $\check{\Pi}$,
the momentum $\check{\mathpzc{p}}$
and the energy flux $\check{\epsilon}$.
The identity \eqref{eq:energy_conservation_z=2_quadr} induces certain
relations between various correlators,
\beq
\langle \check{\mathpzc{p}} \, \check{\mathpzc{p}} \rangle
	=		\lambda \langle \check{\Pi} \, \check{\mathpzc{p}} \rangle
	 =		\lambda^2 \langle \check{\Pi} \, \check{\Pi} \rangle	\; ,	\quad
\langle \check{\mathpzc{p}} \, \check{\epsilon} \rangle
	=		\lambda \langle \check{\Pi} \, \check{\epsilon} \rangle	\; ,
\eeq
leaving three independent two-point correlation functions.
These can be expressed via derivatives of (\ref{eq:v_check},\ref{eq:u_check})
with respect to the sources
$\check{h}^{(0)}$, $\check{\Sigma}^{(0)}$ and $\check{\Sigma}^{(2)}$. 
Up to contact terms,
\bea
\hspace{-7mm}\langle \check{\mathpzc{p}} \, \check{\mathpzc{p}} \rangle &=&
	\frac{ i \omega^2 \lambda^2 }{32 (2\pi)^3 \kappa}
		\left[
				\psi_1
				+ 4 \ln\left(- i \omega\right)
		\right]	\; ,	\label{eq:corr_chp-chp}	\\
\hspace{-7mm}\langle \check{\mathpzc{p}} \, \check{\epsilon} \rangle &=&
	- \frac{ i \omega^3 \lambda^2	\psi_2 }{ 32 (2\pi)^3 \kappa}	\; ,	\label{eq:corr_chp-che}	\\
\hspace{-7mm}\langle \check{\epsilon} \, \check{\epsilon} \rangle &=&
	- \frac{i \omega^4 \lambda }{512 (2\pi)^3 \kappa}
		\left[
			\lambda ( \lambda^2 + 4 ) \psi_1
			- 8 ( \lambda^2 + 2 ) \psi_2
			+ 4 \lambda ( \lambda^2 + 4 ) \ln\left(-i \omega\right)
		\right]		\; .	\label{eq:corr_che-che}
\eea
These correlators can only have poles where $\psi_{1,2}$ have,
that is for $\lambda = 2 i n$, $n \geq 2$.
This implies that all quasinormal frequencies of the shear channel are on the negative
imaginary half-axis and fluctuations in these modes do not cause instabilities.

\section{Sound Channel}
\label{sec:sound}

To construct a general solution for the sound modes \eqref{eq:sound_modes},
it is convenient to introduce
the master field $\mathpzc{Z} = \Omega - 2 a$, which can be shown to satisfy a sixth order
differential equation,
\bea
\mathfrak{D}^3 \frac{1}{v} \mathfrak{D}^2 \frac{1}{v^2} \mathfrak{D}^1 \mathpzc{Z} &=& 0	\; ,
\label{eq:master_eq}
\eea
which used the substitutions \eqref{eq:rTOv}.
The $\mathfrak{D}^j$ denote second order differential operators,
\bea
\mathfrak{D}^1	&=&	v^2 \partial_v^2 + v \left(v-\frac{\lambda}{2}\right)	\; ,	\label{eq:D_op_1}	\\
\mathfrak{D}^2	&=&	v^2 \partial_v^2 + v \partial_v + v \left(v+2 i-\frac{\lambda}{2}\right)	\; ,	\\
\mathfrak{D}^3	&=&	v^2 \partial_v^2 + v \partial_v + v \left(v-2 i-\frac{\lambda}{2}\right)	\; .	\label{eq:D_op_3}
\eea
A general analytic solution for \eqref{eq:master_eq} can be found by introducing functions $\Upsilon^j$ which parameterize the kernels of the operators (\ref{eq:D_op_1}-\ref{eq:D_op_3}), i.e.~$\mathfrak{D}^j \Upsilon^j = 0$.
Explicitly,\footnote{note that $\Upsilon^1$ is of the same form as $\mathpzc{X}$}
\bea
\Upsilon^1
  &=&	A_1 e^{i v}\,\Gamma \left(1+\frac{i\lambda}{4}\right) U\left(\frac{i \lambda}{4} ,0,- 2 i v\right)
  		+ B_1 e^{- i v}\,\Gamma \left(1-\frac{i\lambda}{4}\right) U\left(-\frac{i \lambda}{4} ,0, 2 i v\right)	\; , \quad	\label{eq:Upsilon_1}	\\
\Upsilon^2
  &=&	A_2 e^{i v}\, \Gamma \left(\frac{6 + i \lambda}{4} \right) U\left(\frac{6 + i \lambda}{4} ,1,- 2 i v\right)
  		+ B_2 e^{- i v}\, \Gamma \left(\frac{-2 - i \lambda}{4} \right) U\left(\frac{-2-i \lambda}{4} ,1,2 i v\right)	\; , \quad	\label{eq:Upsilon_2}	\\
\Upsilon^3
  &=&	A_3 e^{i v}\, \Gamma \left(\frac{-2 + i \lambda}{4} \right) U\left(\frac{-2+i \lambda}{4} ,1,- 2 i v\right)
  		+ B_3 e^{- i v}\, \Gamma \left(\frac{6 - i \lambda}{4} \right) U\left(\frac{6 - i \lambda}{4} ,1,2 i v\right)	\; . \quad	\label{eq:Upsilon_3}
\eea
Again, imposing that viable solutions must satisfy
ingoing boundary conditions in the interior implies
$A_j = 0$.
By means of (\ref{eq:Upsilon_1}-\ref{eq:Upsilon_3})
a solution for $\mathpzc{Z}$ can then be constructed explicitly
and from there it is straightforward to successively
find expressions for the other functions in the sound channel.
Summarized, they are all of the form,
\beq
q_0(v) + q_1(v) \Upsilon^j(v) + q_2(v) {\Upsilon^j}'(v) + q_3(v) \mathpzc{I}^j(v) \; ,
\eeq
with
\bea
\mathpzc{I}^j(v) &=& \int_v^\infty \left[ p_1(w) \Upsilon^j(w) + p_2(w) {\Upsilon^j}'(w) \right] \dd  w	\; ,
\eea
and polynomial expressions $q_j$ and $p_j$.
For explicit formul\ae~is referred to app.~\ref{app:explicit_dependence_on_Upsilon_j}.
From these results follows a parameterization for the boundary values of the bulk fields,
\beq
\hspace{-7mm}\Omega^{(0)} =
	\left[\psi_3 + \lambda \ln\left(-i \omega\right)\right] \upsilon_1	
	+ \left[ \psi_4 +(\lambda^2 + 4) \ln\left(-i \omega\right) \right] \upsilon_2
	+ 2 \lambda^2 \upsilon_3
	- 2 (\lambda^2 - 4) \upsilon_4
	- 8 \upsilon_5
	- \upsilon_6	\; ,	\nonumber
\eeq
\beq
\wt{\Omega}^{(0)} =
  \lambda \upsilon_1 + (\lambda^2 + 4) \upsilon_2	\;  ,	\quad
\mathrm{tr}\,h^{(0)} =
	(\lambda^2 + 4)\upsilon_3	\; ,	\quad
\hat{h}^{(0)} =
	(\lambda^2 + 4)\upsilon_4	\; ,	\nonumber
\eeq
\beq
\hat{\Sigma}^{(2)} =
	\omega (\lambda^2 + 4) \upsilon_5	\; ,	\quad
\hat{\Sigma}^{(0)} =
	\lambda \upsilon_6	\; ,	\label{eq:series_coeffs_sound}
\eeq
and their conjugate momenta,
\bea
\hspace{-7mm}T &=&
\frac{ \omega^2 \left[
				3 \lambda (\lambda^2 + 4) 
				- \lambda^2 \psi_3
				- \lambda^3 \ln\left(-i \omega\right)
			\right] }{64} \upsilon_1 \nonumber	\\
				& & \qquad	
	+ \frac{ \omega^2 \left[
				3 (\lambda^2 + 4)^2
				- (\lambda^2 - 4) \psi_4
				- (\lambda^4 - 16)  \ln\left(-i \omega\right)
			\right] }{64} \upsilon_2 \nonumber	\\
				& & \qquad
	+ \frac{ \omega^2  (\lambda^2 + 2) }{2} \upsilon_3
	- \frac{ \omega^2 \lambda^2 }{2} \upsilon_4
	- \omega^2 \upsilon_5
		\; ,	\label{eq:coeff_T}	\\
\hspace{-7mm}\wt{T} &=&
	\frac{\omega^2 \lambda \left( \lambda^2 + 8 \right)}{64} \upsilon_1
	+ \frac{\omega^2 \left( \lambda^2 + 4 \right)^2 }{64} \upsilon_2
		\; , \label{eq:coeff_Tlog}	\\
		& &	\nonumber\\
		& &	\nonumber
\eea
\bea
\hspace{-7mm}\hat{\mathpzc{v}} &=&
	\frac{ \omega \left( \lambda^2 + 4 \right) \left[
				7 \lambda
				- 4 \psi_3
				- 4 \lambda \ln\left(-2 i \omega\right)
			\right] }{64} \upsilon_1
	+ \frac{ \omega \left[
				7 \lambda^2 - 4
				- 4 \psi_4
				- 4 (\lambda^2 + 4) \lambda \ln\left(-i \omega\right)
			\right] }{64} \upsilon_2	\nonumber	\\
				& & \qquad
	+ \frac{ \omega \left[
				- 2 (3 \lambda^2 + 8)
				+ \psi_4
				+ (\lambda^2 + 4)\ln\left(-i \omega\right)
			\right] }{16} \upsilon_3	\nonumber	\\
				& & \qquad
	-\frac{ \omega \left[
				(3\lambda^2 + 4)
				- \psi_4
				- (\lambda^2 + 4)\ln\left(-i \omega\right)
			\right] }{8} \upsilon_4	
	+\frac{ \omega \left[
				8
				+ \psi_4
				+ (\lambda^2 + 4)\ln\left(-i \omega\right)
			\right] }{16} \upsilon_5
		\; .	\label{eq:Vhat}
\eea
The $\upsilon_j$ are just auxiliary variables to parameterize degrees of freedom, and, again,
the $\psi_j$ stand for expressions that involve the Euler--Mascheroni constant
and the digamma function,
\bea
\psi_3	&=&	\frac{1}{2} \left[
					- 3 \lambda + 4 \gamma \lambda - 4 i
					+ \lambda \psi\left(\frac{i \lambda}{4}\right)
				\right]	\; ,	\label{eq:psi_3}	\\
\psi_4	&=&	\frac{1}{2} \left[
					- 3 \lambda^2 + 4 + 4 \gamma (\lambda^2 + 4)
					+ (\lambda^2 + 4)\psi\left(\frac{i \lambda + 2}{4}\right)
				\right]	
	\; .	\label{eq:psi_4}
\eea
The operators sourced by \eqref{eq:series_coeffs_sound}
are the energy $\mathpzc{E}$, an other scalar operator
$\wt{\mathpzc{E}}$ that will be identified as the logarithmic partner of the energy,
as well as
the longitudinal components of
the stress $\hat{\Pi}$,
the momentum $\hat{\mathpzc{p}}$
and the energy flux $\hat{\epsilon}$.
Due to the conservation equation \eqref{eq:momentum_conservation_z=2} 
and the trace condition \eqref{eq:trace_cond_z=2}
the correlators of these operators satisfy a series of identities,
\bea
\langle \mathpzc{E} \, \mathpzc{E} \rangle
	&=&		- \frac{1}{2} \langle \mathpzc{E} \, \mathrm{tr}\Pi\ \rangle
	 =		\frac{1}{\omega} \langle \hat{\epsilon} \, \mathpzc{E} \rangle
	 =		\frac{1}{4} \langle \mathrm{tr}\Pi\, \mathrm{tr}\Pi \rangle
	 =		- \frac{1}{2\omega} \langle \hat{\epsilon} \,\mathrm{tr}\Pi \rangle
	 =		\frac{1}{\omega^2} \langle \hat{\epsilon} \, \hat{\epsilon} \rangle	\;	,\\
\langle \wt{\mathpzc{E}} \, \mathpzc{E}  \rangle
	&=&		- \frac{1}{2} \langle \wt{\mathpzc{E}} \, \mathrm{tr}\Pi \rangle
	 = 		\frac{1}{\omega} \langle \wt{\mathpzc{E}} \, \hat{\epsilon} \rangle	\; ,	\\
\langle \hat{\mathpzc{p}} \, \mathpzc{E} \rangle
	&=&		- \frac{1}{2} \langle \hat{\mathpzc{p}} \, \mathrm{tr}\Pi \rangle
	 =		\frac{1}{\omega} \langle \hat{\mathpzc{p}} \, \hat{\epsilon} \rangle
	 =		\lambda \langle \hat{\Pi} \, \mathpzc{E} \rangle
	 =		- \frac{\lambda}{2} \langle \hat{\Pi} \, \mathrm{tr}\Pi \rangle
	 =		\frac{\lambda}{\omega} \langle \hat{\Pi} \, \hat{\epsilon} \rangle		\; ,	\\
\langle \hat{\mathpzc{p}} \, \wt{\mathpzc{E}} \rangle
	&=&		\lambda \langle \hat{\Pi} \, \wt{\mathpzc{E}} \rangle	\; ,	\\
\langle \hat{\mathpzc{p}} \, \hat{\mathpzc{p}} \rangle
	&=&		\lambda \langle \hat{\Pi} \, \hat{\mathpzc{p}} \rangle
	 =		\lambda^2 \langle \hat{\Pi} \, \hat{\Pi} \rangle	\; .
\eea
This leaves six independent two-point functions. 
Extracting these from the on-shell solution is a bit more involved than it was in
the shear channel, but despite a little bit of algebra to invert
\eqref{eq:series_coeffs_sound}
it is again rather straightforward,
\bea
\hspace{-7mm}\langle \mathpzc{E} \, \mathpzc{E} \rangle
  &=&	-\frac{i \omega^2 \lambda (\lambda^2+4)}{8 \kappa (2\pi)^3 \Phi}	\; ,	\label{eq:corr_EE_Fourier}	\\
\hspace{-7mm}\langle \wt{\mathpzc{E}} \mathpzc{E} \rangle
  &=&	\frac{i \omega^2 \lambda (\lambda^2+4)}{8 \kappa (2\pi)^3 \Phi} \ln\left(-i \omega\right)
  		+\frac{i \omega^2	\left[
  									(\lambda^2+4)\psi_3 + \lambda \psi_4	
  							\right]}{16 \kappa (2\pi)^3 \Phi}	\; ,	\label{eq:corr_EElog_Fourier}	\\
\hspace{-7mm}\langle \wt{\mathpzc{E}} \wt{\mathpzc{E}} \rangle
  &=&	- \frac{i \omega^2 \lambda (\lambda^2+4)}{8 \kappa (2\pi)^3 \Phi} \ln\left(-i \omega\right)^2
  		- \frac{i \omega^2	\left[
  									(\lambda^2+4)\psi_3 + \lambda \psi_4	
  							\right]}{8 \kappa (2\pi)^3 \Phi} \ln\left(-i \omega\right)
  		- \frac{i \omega^2	\psi_3 \psi_4 }{8 \kappa (2\pi)^3 \Phi} 	\; ,	\label{eq:corr_ElogElog_Fourier}	\\
\hspace{-7mm}\langle \hat{\mathpzc{p}} \, \mathpzc{E} \rangle &=&
	\frac{i \omega^2\lambda^2}{\kappa (2\pi)^3 \Phi }	\; ,	\label{eq:corr_pE}	\\
\hspace{-7mm}\langle \hat{\mathpzc{p}} \, \wt{\mathpzc{E}} \rangle &=&
	- \frac{ 2 i \omega^2\lambda \left[ \psi_3 + \lambda \ln\left(-i \omega\right) \right]}{\kappa (2\pi)^3 \Phi }	\; ,	\label{eq:corr_pElog}	\\
\hspace{-7mm}\langle \hat{\mathpzc{p}} \, \hat{\mathpzc{p}} \rangle &=&
	- \frac{i \omega^2 \lambda^2 }{8 \kappa (2\pi)^3 (\lambda^2+4)}
		\left[
			\frac{64 \lambda}{\Phi}
			-\psi_4
			-(\lambda^2+4) \ln\left(-i \omega\right)
		\right]		\; .	\label{eq:corr_pp}
\eea
The function $\Phi$ appearing in the denominator of the expressions above is related to 
the determinant of the set of linear equations \eqref{eq:series_coeffs_sound}
and can be expressed in closed form,
\bea
\Phi &=&
	2 i \lambda^2 +8 \lambda +8 i 
	- \left(\lambda^2+4\right) \lambda  \psi\left(\frac{i \lambda }{4}\right)
	+ \left(\lambda^2+4\right) \lambda  \psi\left(\frac{i \lambda + 2}{4}\right)	\; .
\label{eq:Phi}
\eea
Though it may not appear so at first sight, 
the correlator in \eqref{eq:corr_pp} is actually regular at $\lambda = -2 i$.
It easily can be verified that the residue at that point vanishes due to
$\psi_4(-2i) = 8$ and $\Phi(-2i) = -16i$.
Furthermore, $\Phi$ cancels the poles coming from the numerator.
Thus, with the exception of
$\langle \hat{\mathpzc{p}} \, \hat{\mathpzc{p}} \rangle$
that also contains poles at $\lambda = 2 (2n+1)i$, $n \geq 1$,
all quasinormal modes are determined by the
zeros of \eqref{eq:Phi}.
A closed expression for those values is not at hand, but it is possible to argue
stability in the sound channel by showing that these zeros
must all be located in the upper half of the complex $\lambda$-plane.
For this purpose, define,
\beq
\xi(\mu,\nu)	=	\Re \Phi(\mu+i \nu)	\; , \quad
\eta(\mu,\nu)	=	\Im \Phi(\mu+i \nu)	\; .
\eeq
The zeros of $\Phi$ then correspond to the intersections of the set of
curves defined by $\xi = 0$ and $\eta = 0$.
The non-existence of zeros in the lower half-plane follows if it can be shown that
$\eta$ is nowhere zeros in that region.
This will be done by excluding all other possibilities.
As $\eta$ is a harmonic function that asymptotes to $6\nu\left(1 + \frac{2}{\mu^2 + \nu^2}\right)$ 
for large $|\mu|$ and $|\nu|$, the
set $\eta = 0$ in the lower half-plane must be a collection of smooth curves, each of which
must be in one of three possible categories,
\begin{enumerate}[(i)]
\item
a bounded closed curve in $\R_{<0}\times\R$,
\item
a bounded curve that intersects with the real axis $\nu=0$,
\item
a curve that asymptotes to the real axis $\nu=0$ for large $|\nu|$.
\end{enumerate}
Case (i) would imply that the curve would encircle a pole.
This is impossible as all poles of $\Phi$ are in the upper half-plane.
Case (ii) can be excluded by examining $\eta$ on the real axis,
\beq
\eta(\mu,0)
	=-\frac{\pi \mu\, (4+\mu^2)}{\mathrm{Sinh}\left( \frac{\pi \mu}{2} \right)}
	\label{eq:eta_nu0}	\; ,
\eeq
which is strictly negative.
Finally, to exclude case (iii), consider
\bea
\frac{\partial\eta(\mu,0)}{\partial \nu} 
	&\;\;\stackrel{\mu \to \pm \infty}{\xrightarrow{\hspace*{11mm}}}\;\;&	6 + \frac{12}{\mu^2}	\; .
\eea
Together with $\eta(\mu,0)$ being strictly negative on the real axis,
this implies that the curve that asymptotes to the real axis must lie in the
upper half-plane.
This finishes the proof that $\Phi$ has no zeros with negative imaginary part.

\begin{figure}[t]
\begin{center}
\includegraphics[height=.5\paperwidth]{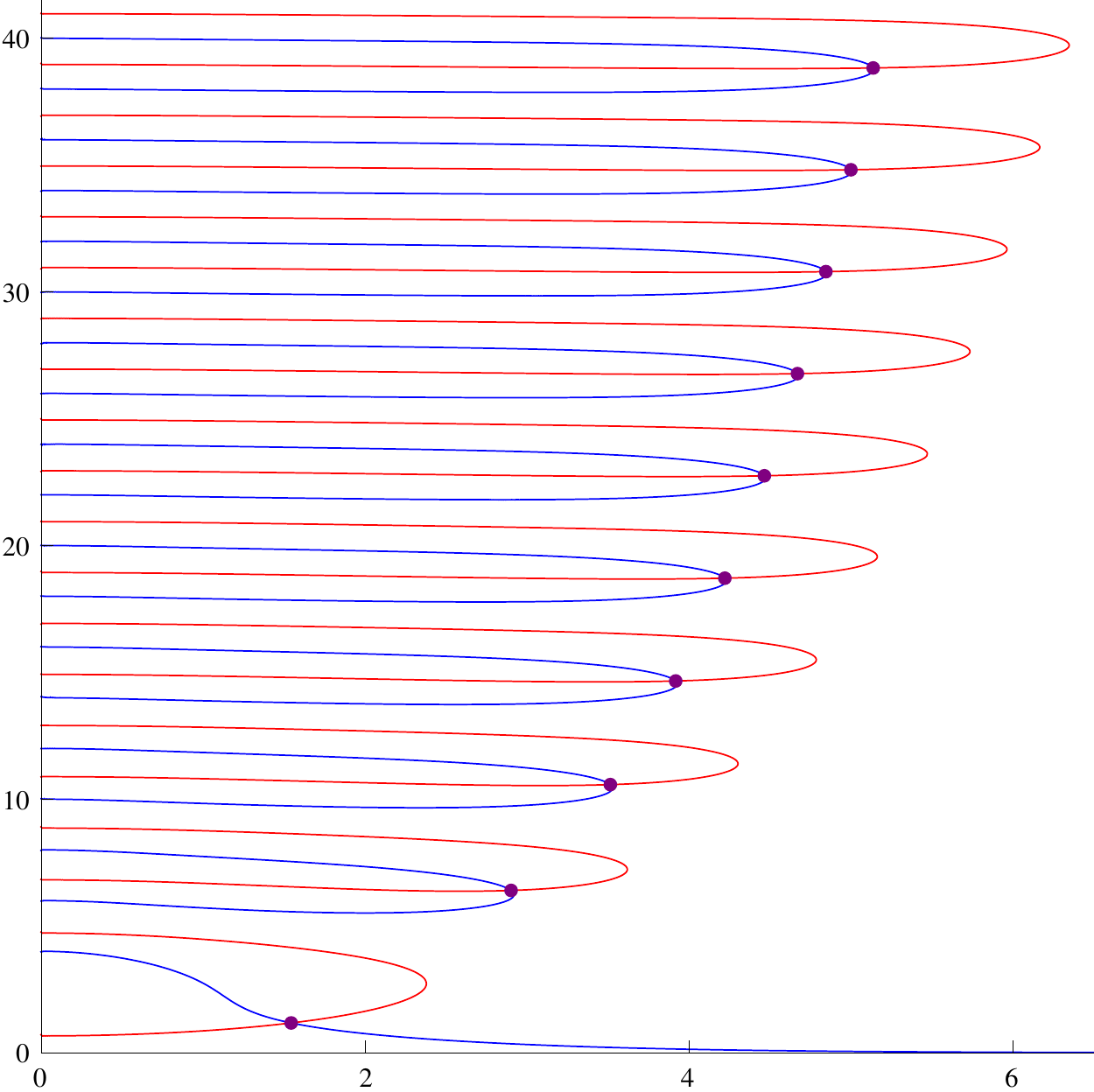}
\caption{\label{fig:zeros} The zeros of $\Phi$ in the right upper half-plane.
Red curves mark the vanishing of the real part, blue curves the vanishing of the imaginary part.
The intersections, marked by the purple dots, are the zeros.}
\end{center}
\end{figure}
Analyzing the positions of zeros in the upper half-plane is more complicated.
What can be shown in a straightforward way is that there can be no zeros on the
imaginary axis.
The condition $\Phi(0,\nu) = 0$ is equivalent to finding real solutions of
\bea
\frac{2\pi \nu\, (\nu - 2)}{\sin\left(\frac{\pi \nu}{2}\right)}
	&=& \nu\, (\nu - 2)\left[\psi\left(\frac{\nu + 2}{4}\right) - \psi\left(\frac{\nu + 4}{4}\right)\right]
			+\frac{2 \,(\nu^2 - 4 \nu - 4)}{( \nu + 2 )}	\; .	\label{eq:nu_eq_0}
\eea
For positive $\nu$, the right hand side is a negative function 
bounded from below by $-6$,
whereas the minimum of the absolute value of the left hand side 
in this range is $2 \pi$.
Hence, \eqref{eq:nu_eq_0} can have no real solutions.
There are however zeros of $\Phi$ away from the imaginary axis.
In a similar fashion as before, it is possible to analyze
what restrictions the poles and extrema on the imaginary axis
impose on the curves $\xi = 0$ and $\eta = 0$.
The conclusion is that there must be an infinite set of intersections of these curves
away from the imaginary axis and the zeros of $\Phi$ therefore
consist of an infinite set of pairs $\{\lambda, -\bar{\lambda} \}$.
For large $n \in \N$, there must be one such pair
with $4n > \Im \lambda > 4n-2$, but 
finding exact values for zeros was
beyond the scope of analytic methods
and it was necessary to resort to numerics,
a plot is shown in fig.~\ref{fig:zeros}.

\subsection{Logarithmic structure in two-point correlation functions}
\label{sec:log}

This section deals with $\mathpzc{E}$ and $\wt{\mathpzc{E}}$, i.e.~the operators 
sourced by $\Omega^{(0)}$ and $\wt{\Omega}^{(0)}$ respectively.
The aim is to identify $\wt{\mathpzc{E}}$ as a logarithmic partner of the energy operator $\mathpzc{E}$
by comparing their two-point functions to (\ref{eq:LCFT_2pt_1}-\ref{eq:LCFT_2pt_3}),
that is to the structure of increasing powers of logarithmic terms appearing
when operators belong to an indecomposable non-diagonalizable representation.
The correlators (\ref{eq:corr_EE_Fourier}-\ref{eq:corr_ElogElog_Fourier}) are
already quite suggestive about the existence of such a tower of logarithmic terms,
though in order to have a more elementary comparison it would be desirable to have an
expression in position space.
Unfortunately, given the complicated form of the denominator, translating
the momentum space correlators  back to position space in all generality seems out of question.
Though, due to a general knowledge of the position of poles and other
features of the functions, it is
still possible to qualitatively deduce the underlying structure of
two-point correlation functions,
\bea
\langle \mathpzc{E}(t,x) \, \mathpzc{E}(0,0) \rangle
  &=&	\frac{\theta(t)}{t^4} \,\mathpzc{g}^{(0)}(\chi)		\; ,	\label{eq:corr_EE}	\\
\langle \wt{\mathpzc{E}}(t,x) \, \mathpzc{E}(0,0) \rangle
  &=&	\frac{\theta(t) \ln t}{t^4} \,\mathpzc{g}^{(0)}(\chi)
  		+\frac{\theta(t)}{t^4} \,\mathpzc{g}^{(1)}(\chi)	\; ,	\label{eq:corr_EElog}	\\
\langle \wt{\mathpzc{E}}(t,x) \, \wt{\mathpzc{E}}(0,0) \rangle
  &=&	\frac{\theta(t)\ln^2 t}{t^4} \,\mathpzc{g}^{(0)}(\chi)
  		+\frac{\theta(t)\ln t}{t^4} \,\mathpzc{g}^{(2)}(\chi)
  		+\frac{\theta(t)}{t^4} \,\mathpzc{g}^{(3)}(\chi)	\; .	\label{eq:corr_ElogElog}
\eea
The functions $\mathpzc{g}^{(j)}$ solely depend
on the ratio $\chi = x^2/ 4 t$, as is to be expected
for a theory with $z = 2$ Lifshitz scaling.
Further details about the transformation back to position space can be found in
app.~\ref{app:Fourier}.
The structure of these two-point functions bears a resemblance to the logarithmic
field theory with Lifshitz scaling developed in~\cite{Bergshoeff:2011xy}.
The power in the prefactor suggests that $\mathpzc{E}$ and $\wt{\mathpzc{E}}$
are operators with dimension $2$ with respect to $t$, respectively $4$ with respect to $x$,
consistent with the interpretation of energy in a theory with Lifshitz scaling.
The mechanism behind the appearance of this
logarithmic pair can be argued as follows.
For general dynamical exponent $z$ a Lifshitz fixed point can occur when there is
a certain relation between
the timelike component of the tetrad $e^0$ and the Proca field $P$.
The modes of these two fields combine to form sources for a set of
operators in the dual field theory\footnote{cf sec.~\ref{sec:asylif}} --
in particular $\mathpzc{E}$ and $\mathpzc{O}_P$ --
which, generically,
would be part of a diagonalizable representation of the underlying algebra.
The limit $z \to 2$ induces a degeneracy that requires to introduce Jordan cells
to get a complete description
and, taking CFT as a guideline, it is exactly the appearance of such cells that
would make the field theory logarithmic.

The most striking difference is the presence of $\mathpzc{g}^{(0)}$,
i.e.~that when comparing to the LCFT correlators~(\ref{eq:LCFT_2pt_1}-\ref{eq:LCFT_2pt_3})
one would expect $\mathpzc{g}^{(0)} = 0$.
This term did also not appear in the toy model in~\cite{Bergshoeff:2011xy}.
However, the vanishing of the right hand side of
\eqref{eq:LCFT_2pt_1} is only stringent for a LCFT with
proper primaries and
it thus could have a non-zero value if the OPE of two primaries
contains logarithmic fields.
Beyond that, other conditions that require
the vanishing of this two-point function are related to 
the underlying conformal symmetry.
Therefore it stands to reason that requiring a less stringent level of symmetry
-- like anisotropic scaling -- would also be less restrictive about the values
of certain coefficients in the correlation functions.

The function $\mathpzc{g}^{(0)}$ also distinguishes itself from
$\mathpzc{g}^{(1,2,3)}$ in that it for $t \ll x^2$,
respectively  $\chi \gg 1$,
incorporates a certain ultralocal and oscillatory behavior,
\bea
\hspace{-7mm} \mathpzc{g}^{(0)}	& \propto &
	\chi^3 e^{- \Im \lambda_{\min} \chi}
		\Re\left[ \mathpzc{A}\,  L_{3}\left( -i\lambda_{\min}\chi \right) e^{i \Re \lambda_{\min} \chi} \right] \; ,
\eea
where $\mathpzc{A}$ is an amplitude that depends on $\lambda_{\min}$,
the latter being a zero of $\Phi$ with minimal imaginary part.
The other functions are only expected to follow a power law
in this limit, such that for fixed $t > 0$ and $x^2 \gg t$,
\bea
\langle \mathpzc{E}(t,x) \, \mathpzc{E}(0,0) \rangle 
	&\stackrel{x^2 \gg t}{\xrightarrow{\hspace*{8mm}}}&
	0	\; ,	\label{eq:corr_EE_large_chi}	\\
\langle \wt{\mathpzc{E}}(t,x) \, \mathpzc{E}(0,0) \rangle
	&\stackrel{x^2 \gg t}{\xrightarrow{\hspace*{8mm}}}&
		\frac{A^{(1)}}{x_4}	\; ,	\label{eq:corr_EElog_large_chi}	\\
\langle \wt{\mathpzc{E}}(t,x) \, \wt{\mathpzc{E}}(0,0) \rangle
	&\stackrel{x^2 \gg t}{\xrightarrow{\hspace*{8mm}}}&
		\frac{A^{(2)}\ln x}{x_4}
  		+\frac{A^{(3)}}{x_4}		\; ,	\label{eq:corr_ElogElog_large_chi}
\eea
with some constants $A^{(j)}$.
Interestingly, up to normalization, this agrees well with
(\ref{eq:LCFT_2pt_1}-\ref{eq:LCFT_2pt_3}),
i.e.~the expected form of two-point functions in a LCFT
and can be seen as a further indication that the underlying
structure in the correlators (\ref{eq:corr_EE}-\ref{eq:corr_ElogElog}) is indeed
an extension of the concept of logarithmic field theories to the case
of anisotropic scaling symmetry.
In the opposite limit, i.e.~fixed $x$ and $t \gg x^2$,
the functions $\mathpzc{g}^{(j)}$ can be expanded in a power series.
Including terms up to first order,
\bea
\langle \mathpzc{E}(t,x) \, \mathpzc{E}(0,0) \rangle 
	&\stackrel{t \gg x^2}{\xrightarrow{\hspace*{8mm}}}&
	\frac{\chi\, C^{(0)} }{t^4}		\; ,	\label{eq:corr_EE_small_chi}	\\
\langle \wt{\mathpzc{E}}(t,x) \, \mathpzc{E}(0,0) \rangle
	&\stackrel{t \gg x^2}{\xrightarrow{\hspace*{8mm}}}&
		\frac{\chi\, C^{(0)} \ln t}{t^4}
  		+\frac{\left( B^{(i)} + \chi\, C^{(i)} \right)}{t^4}	\; ,	\label{eq:corr_EElog_small_chi}	\\
\langle \wt{\mathpzc{E}}(t,x) \, \wt{\mathpzc{E}}(0,0) \rangle
	&\stackrel{t \gg x^2}{\xrightarrow{\hspace*{8mm}}}&
		\frac{\chi\, C^{(0)}  \ln^2 t}{t^4} \,
  		+\frac{\left( B^{(ii)} + \chi\, C^{(ii)} \right)\ln t}{t^4} \,
  		+\frac{\left( B^{(iii)} + \chi\, C^{(iii)} \right)}{t^4}	\; ,	\label{eq:corr_ElogElog_small_chi}
\eea
with constants $B^{(j)}$ and $C^{(j)}$.
Also here, the leading logarithmic term is suppressed, though not exponentially as in the case before.

\section{Summary and Outlook}

This paper dealt with an investigation of perturbations around an asymptotically
$z=2$ Lifshitz fixed point of the Einstein--Proca action. 
The main focus was on calculating two-point functions and what information can
be gained from this about the dual Lifshitz field theory.
It was found that a degeneracy which resulted in the appearance of logarithmic modes
in the Fefferman--Graham expansion also lead to logarithmic terms 
in two-point correlation functions in Fourier space.
When transforming back to position space,
correlators containing these terms fail to be ultralocal
in the limit $x^2 \gg t$, which otherwise seems the generic behavior.
A second, and more interesting, feature caused by those terms
was the structure of correlation functions involving the energy
$\mathpzc{E}$ and a second operator $\wt{\mathpzc{E}}$.
The latter was identified as a logarithmic partner of the former, due to
the similarities of their correlators to properties
of LCFTs and a toy model for a logarithmic field theory with Lifshitz scaling.
This lead to the conclusion that $z=2$ Lifshitz solutions
of the Einstein--Proca model are candidates for gravitational duals
of logarithmic Lifshitz theories.

What further could be extracted from the investigation of two-point functions
was the position of quasinormal modes in the complex frequency plane.
In the shear channel these can be calculated explicitly and are all
found to lie on the negative imaginary axis.
The sound channel also contains quasinormal modes
on the negative imaginary axis, but, in addition to that,
also a set of modes that come in pairs $\{\omega, -\bar{\omega}\}$
with non-vanishing real part.
For the latter, analytic values could not be obtained,
but it was possible to prove that they are all
located in the lower half-plane of complex frequencies.
This indicates the stability of fluctuations around this Lifshitz
spacetime -- at least at linearized level.

The model considered in this paper was a specific $z=2$ Lifshitz theory
in $2+1$ dimensions, leaving the question of how results could be generalized.
Logarithmic modes are certainly a rather specific feature of the case
at hand and can not be expected to appear for generic $z$.
Considering a $z=n$ Lifshitz theory in $n+1$ dimensions
seems straightforward, given that it indeed contains similar degeneracies
in the Fefferman--Graham expansion as the $z=2$ case.
However, explicit results could not be obtained directly
as an analytic solution to the equations of motion was not found.
It would nonetheless be possible to search for the appearance of logarithmic
correlators by numerical analysis, but this will be left for future study.

An other possible direction to go from the results obtained here would
be the analysis of three-point functions and other higher order correlators.
Though it would in principle be straightforward to get analytic results,
obtaining them will likely require a certain level of tenaciousness and endurance,
if not a more suitable formalism for these kind of calculations,
as the equations of motion and the renormalized action tend
to become rather cumbersome when higher order expansions are considered.
Nevertheless, it would be interesting to see whether the similarities 
between LCFTs and the model considered here persist to that level.

\section*{Acknowledgements}

This work was supported by the Nederlandse Organisatie voor Wetenschappelijk Onderzoek (NWO) under the research program of the Stichting voor Fundamenteel Onderzoek der Materie (FOM).

\newpage

\appendix

\section{Counterterms}
\label{app:cts}

In order to write down the counterterms, the boundary is written as $\R \times \mathpzc{C}$ with a $2$-manifold ${\mathpzc{C}}$, i.e.~split into timelike and spacelike parts.
Using this decomposition,
\beq
\bar{e}^0 = e^0\Bigr|_{\mathpzc{C}^{\perp}} \; , \quad
\bar{\Sigma} = e^0\Bigr|_{\mathpzc{C}} \; ,	\quad
\bar{e}^j = e^j\Bigr|_{\mathpzc{C}} \; \quad
\bar{\mathpzc{a}} = P - \Sigma\,\Bigr|_{\mathpzc{C}^{\perp}} \; , \quad
\bar{\mathpzc{b}} = P - \Sigma\,\Bigr|_{\mathpzc{C}} \; .
\eeq
A differential $\bar{\dd}$ on $\mathpzc{C}$ is also naturally understood
via restriction of $\dd$ to ${\mathpzc{C}}$.
Furthermore, a Hodge star operator on $\mathpzc{C}$ can be defined,
\beq
\bar{*} 1 = \bar{e}^1 \wedge \bar{e}^2 = \bar{v}  \; , \quad
\bar{*} \bar{e}^1 = \bar{e}^2 \; , \quad
\bar{*} \bar{e}^2 = -\bar{e}^1 \; .
\eeq
This allows to define codifferential $\bar{\delta}$ and Laplacian $\bar{\Delta}$
in a standard way.
In order to construct terms which are invariant under a rotation 
of $\{ \bar{e}^1, \bar{e}^2 \}$, define for $\xi$, a general vector field,
\bea
\mathfrak{T}_{\xi} \bar{e}^1
	&=&	\mathfrak{L}_{\xi} \bar{e}^1
			+ \frac{1}{2} \left[
					\bar{e}^{1}\formprod \mathfrak{L}_{\xi} \bar{e}^2
				- \bar{e}^2\formprod \mathfrak{L}_{\xi} \bar{e}^1 \right] \bar{e}^2	\; ,	\\
\mathfrak{T}_{\xi} \bar{e}^2
	&=&	\mathfrak{L}_{\xi} \bar{e}^2
			+ \frac{1}{2} \left[
					\bar{e}^2 \formprod \mathfrak{L}_{\xi} \bar{e}^1
					- \bar{e}^1 \formprod \mathfrak{L}_{\xi} \bar{e}^2 \right] \bar{e}^1	\; .
\eea
Furthermore, in the following the notation is used that the action of spacelike derivative
operations on the
timelike components is to be understood as the action on a $0$-form on ${\mathpzc{C}}$.
Timelike derivatives can be generated by $\mathfrak{L}_{\bar{e}^0}$,
\beq
\dot{\bar{\mathpzc{a}}}
	=	\left( \bar{e}^0 \formprod \mathfrak{L}_{\bar{e}^0} \bar{\mathpzc{a}} \right) \bar{e}^0	\; ,	\quad
\dot{\Sigma}
	=	\LieDer{\bar{e}^0}{\bar{\Sigma}}	\; ,	\quad
\dot{\bar{v}}
	=	\LieDer{\bar{e}^0}{\bar{v}}	\; ,	\nonumber
\eeq
\beq
\dot{\bar{e}}^1
	=	\mathfrak{L}_{\bar{e}^0} \bar{e}^1
			+ \frac{1}{2} \left[
					\bar{e}^{1}\formprod \mathfrak{L}_{\bar{e}^0} \bar{e}^2
				- \bar{e}^2\formprod \mathfrak{L}_{\bar{e}^0} \bar{e}^1 \right] \bar{e}^2	\; ,	\quad
\dot{\bar{e}}^2
	=	\mathfrak{L}_{\bar{e}^0} \bar{e}^2
			+ \frac{1}{2} \left[
					\bar{e}^2 \formprod \mathfrak{L}_{\bar{e}^0} \bar{e}^1
					- \bar{e}^1 \formprod \mathfrak{L}_{\bar{e}^0} \bar{e}^2 \right] \bar{e}^1	\; .
\eeq
With this notation, explicit expressions for counterterms
that make the on-shell action finite,
at quadratic order, are as follows,
\bea
\hspace{-7mm} \mathpzc{S}^{ct}
	&=&	\int
		\bar{e}^0 \wedge \biggl[
			\left(
				K
				+6
				+\bar{R}			
			\right)  \bar{v}	
			-2\, \dot{\bar{e}}^1 \wedge \dot{\bar{e}}^2
			+ \frac{1}{4} \dot{\Sigma} \wedge \bar{*} \dot{\bar{\Sigma}}
			- \bar{\mathpzc{b}} \wedge \bar{*} \bar{\mathpzc{b}}	\nonumber	\\
	& &	\qquad\qquad\qquad
			- \frac{1}{2} \bar{\dd}\bar{\mathpzc{b}} \wedge \bar{*} \bar{\dd}\bar{\Sigma}
			- \frac{1}{2} \bar{\delta} \dot{\bar{\Sigma}} \wedge \bar{v}
			-\bar{\delta} \bar{\mathpzc{b}} \wedge	\dot{\bar{v}}
			+ \dot{\bar{e}}^1 \wedge \mathfrak{T}_{\bar{\Sigma}} \bar{e}^2 
			- \dot{\bar{e}}^2 \wedge \mathfrak{T}_{\bar{\Sigma}} \bar{e}^1
		\biggr]	\nonumber	\\
	& &	\qquad
			-\frac{1}{4} \bar{\dd} \bar{e}^0 \wedge \bar{*} \bar{e}^0 \formprod \bar{\dd} {\bar{e}^0}
	\nonumber	\\
	& &	\qquad + \bar{\mathpzc{a}} \wedge \left[
			\left( 2 - \bar{R} \right) \bar{v}
			+ \ddot{\bar{v}}
			+ \frac{1}{2} \bar{*} \bar{e}^0 \formprod\bar{\Delta} \bar{e}^0
			- 2 \bar{*} \bar{e}^0 \formprod \bar{\mathpzc{a}}
			- \frac{3}{4} \bar{*} \bar{e}^0 \formprod \bar{\Delta}\bar{\mathpzc{a}}
			- \frac{1}{2} \bar{*} \bar{\delta} \dot{\bar{\Sigma}}
		\right]
\; ,	\\
\hspace{-7mm} \mathpzc{S}^{ct,anom}
	&=&	\int \ln r \,	\bar{e}^0 \wedge \biggl[
			2 \, \dot{\bar{e}}^1 \wedge \dot{\bar{e}}^2
			- \frac{1}{2} \dot{\bar{v}} \wedge \bar{*} \dot{\bar{v}}
			- \bar{\delta} \bar{\mathpzc{b}} \wedge	\dot{\bar{v}}	
			- \frac{1}{2} \bar{\mathpzc{b}} \wedge \bar{*} \bar{\Delta}\bar{\mathpzc{b}}\nonumber	\\
	& &	\qquad\qquad\qquad
			+ \frac{1}{8} \bar{\dd}\dot{\bar{\Sigma}} \wedge \bar{*} \bar{\dd}\dot{\bar{\Sigma}}
			+ \frac{1}{4} \bar{\dd} \bar{\mathpzc{b}} \wedge \bar{*} \bar{\dd}\bar{\Delta}\bar{\Sigma}
			+ \frac{1}{2} \bar{R} \, \bar{\delta}\dot{\bar{\Sigma}} \wedge \bar{v}	\nonumber	\\
	& &	\qquad\qquad\qquad
			+ 2 \dot{\bar{e}}^1 \wedge \mathfrak{T}_{\bar{\mathpzc{b}}} \bar{e}^2 
			- 2 \dot{\bar{e}}^2 \wedge \mathfrak{T}_{\bar{\mathpzc{b}}} \bar{e}^1
			- \frac{1}{2} \dot{\bar{e}}^1 \wedge \mathfrak{T}_{\bar{\Delta}\bar{\Sigma}} \bar{e}^2 
			+ \frac{1}{2} \dot{\bar{e}}^2 \wedge \mathfrak{T}_{\bar{\Delta}\bar{\Sigma}} \bar{e}^1
		\biggr]	\;.
\eea

\section{Sound Channel Solutions}
\label{app:explicit_dependence_on_Upsilon_j}

Explicit expression for solutions to the equations
(\ref{eq:lineom_1}-\ref{eq:lineom_10}) and constraints
(\ref{eq:constr_1}-\ref{eq:constr_4}) for the sound
channel modes \eqref{eq:sound_modes} can be split into 
four contributions, $\xi = \xi^0+\xi^1+\xi^2+\xi^3$.
The first one originates from three integration constants
and takes a rather simple form,
\bea
\Omega^0 &=&
	\mathpzc{c}_1 + \mathpzc{c}_2 v^2	\; ,	\\
a^0 &=&
	\frac{\mathpzc{c}_1 + \mathpzc{c}_2 v^2}{2}	\; , \\
\mathrm{tr}\, h^0 &=&
	\mathpzc{c}_3 - \lambda \mathpzc{c}_2 v	\; , \\
\hat{h}^0 &=&
	\mathpzc{c}_3 - \mathpzc{c}_2 - \lambda \mathpzc{c}_2 v	\; , \\
\hat{\Sigma}^0 &=&
	-\lambda ( \mathpzc{c}_1 - 2 \mathpzc{c}_2)
	- 2 (2\mathpzc{c}_2-\mathpzc{c}_3) v
	- 3 \lambda \mathpzc{c}_2 v^2 
	\; , \\
\hat{b}^0 &=&
	-\frac{\lambda}{2} (\mathpzc{c}_1 - 2 \mathpzc{c}_2 + \mathpzc{c}_2 v^2)	\; , \\
c^0	 &=&
	i \mathpzc{c}_2 v	\; .
\eea
The other contribution vanish in the limit
$v \to \infty$ while satisfying infalling boundary conditions.
They come from the functions $\Upsilon^j$, as they are
given by the expressions (\ref{eq:Upsilon_1}-\ref{eq:Upsilon_3}).
In order to proceed, first define,
\bea
\mathpzc{I}^1(v)	&=&
  \int_v^{\infty } \left((4 + 3 i \lambda -8 i w) \Upsilon^1(w)
  + (4 i -\lambda +8 w) {\Upsilon^1}'(w)\right) \, \dd w	\; ,	\\
\mathpzc{I}^2(v)	&=&
  \int_v^{\infty }  \Upsilon^2(w) \, \dd w	\; ,	\\
\mathpzc{I}^3(v)	&=&
  \int_v^{\infty} \left(
  \left(4 + \lambda^2 -16 (\lambda +i) w + 32 w^2\right) \Upsilon^3(w)
  -8 i w (\lambda - 4 w) {\Upsilon^3}'(w)
		  \right) \, \dd w	\; .
\eea
Expanding these into a power series around $v=0$
requires the explicit calculation of the integrals
$\iota^j = \mathpzc{I}^j(0)$.
These can be evaluated by elementary methods,
\bea
\iota^1
    &=&	\frac{(4+i \lambda ) \lambda}{4} \left[\psi\left(\frac{12+i \lambda}{8}\right) - \psi\left(\frac{8+i \lambda}{8}\right) \right]	\; ,	\\
\iota^2
    &=&	\frac{i}{2} \left[\psi\left(\frac{6+i \lambda}{8}\right) - \psi\left(\frac{10 + i \lambda}{8}\right) \right]	\; ,	\\
\iota^3
    &=&	\frac{i(4+\lambda^2)}{2} \left[\psi\left(\frac{6+i \lambda}{8}\right) - \psi\left(\frac{10 + i \lambda}{8}\right) \right]	\; .
\eea
An explicit  general solution for the functions in the sound channel
can now be given
in terms of $\mathpzc{I}^j $, $\Upsilon^j$ and ${\Upsilon^j}'$.

\subsection*{{$\Upsilon^1$} :}
\bea
\hspace{-7mm}\Omega^1 &=&
	\frac{i}{2 \lambda  (4 i - \lambda)} \Bigl[
	\left(8-\lambda^2+4 v^2\right) \mathpzc{I}^1	\nonumber	\\
& &	\qquad +\left( 32 i + 16 \lambda + 6 i \lambda^2 + \lambda^3 -8 \left(8+\lambda^2\right) v -4 (\lambda +4 i) v^2 + 32 v^3 \right) \Upsilon^1 \nonumber	\\
& &	\qquad -2 \left( 16 +4 i \lambda +\left( 32 i -4\lambda +3 i \lambda^2 \right) v - 2(4+i \lambda) v^2 - 16 i v^3 \right) {\Upsilon^1}'
	 \Bigr]	\; ,
\eea
\bea
\hspace{-7mm}a^1 &=&
	\frac{i}{4 \lambda  (4i - \lambda)} \Bigl[
	- \left( 8+\lambda^2-4 v^2\right) \mathpzc{I}^1	\nonumber	\\
& &	\qquad +\left( 32 i +32 \lambda+10 i \lambda^2 +\lambda^3 - 8 \left(8+\lambda^2\right) v - 4 (4 i+\lambda) v^2 +32 v^3 \right) \Upsilon^1	\nonumber	\\
& &	\qquad - 2 \left(16 +4 i \lambda +\left(32 i + 3 i \lambda^2-4 \lambda \right) v-2(4 + i \lambda ) v^2+16 i v^3 \right) {\Upsilon^1}'
	\Bigr]	\; , \\
\hspace{-7mm}\mathrm{tr}\,h^1 &=&
	\frac{i}{\lambda  (4 i - \lambda)} \Bigl[
      \left( 4 + \lambda^2 - 2 \lambda v \right) \mathpzc{I}^1	\nonumber	\\
& &	\qquad - \left( 16 i + 4\lambda +4i\lambda^2+\lambda^3 -2 \left( 16 +4 i \lambda + 5 \lambda^2 \right) v + 16 \lambda v^2 \right) \Upsilon^1	\nonumber	\\
& &	\qquad - 2 i \left(8 i -2 \lambda -\left( 16+4 i \lambda+3 \lambda^2 \right) v+ 8 \lambda  v^2 \right) {\Upsilon^1}'
	\Bigr]	\; ,	\\
\hspace{-7mm}\hat{h}^1 &=&
	\frac{i}{\lambda  (4 i - \lambda)} \Bigl[
      \left( 2 + \lambda^2 - 2 \lambda v \right) \mathpzc{I}^1	\nonumber	\\
& &	\qquad - \left( 8 i -6\lambda +2i\lambda^2+\lambda^3 -2 \left( 8 +4 i \lambda + 5 \lambda^2 \right) v + 16 \lambda v^2 \right) \Upsilon^1	\nonumber	\\
& &	\qquad - 2 i \left(4 i -\lambda -\left( 8+4 i \lambda+3 \lambda^2 \right) v+ 8 \lambda  v^2 \right) {\Upsilon^1}'
	\Bigr]	\; ,	\\
\hspace{-7mm}\hat{\Sigma}^1 &=&
	\frac{i}{2 (4 i - \lambda )}  \Bigl[
	+ \left(16 + \lambda^2 + 4 \lambda v-12 v^2\right) \mathpzc{I}^1	\nonumber	\\
& &	\qquad -\left(64 i+8 \lambda+2 i \lambda^2+\lambda^3 - 4\left(32-4 i \lambda+\lambda^2 \right) v -4 (12 i+11 \lambda) v^2+96 v^3\right) \Upsilon^1	\nonumber	\\
& &	\qquad + 2 v \left(64 i-4 \lambda+3 i \lambda^2 - 2 (12-5 i \lambda ) v - 48 i v^2\right) {\Upsilon^1}'
	\Bigr]	\; ,	\\
\hspace{-7mm}\hat{b}^1 &=&
	\frac{i}{4 (4 i - \lambda)}  \Bigl[
	\left(16+\lambda^2-4 v^2\right) \mathpzc{I}^1	\nonumber	\\
& &	\qquad -\left(64 i+8 \lambda +2 i \lambda^2+\lambda^3 -4 (4 i+\lambda) v^2-8 \left(8-2 i \lambda+\lambda^2\right)v +32 v^3 \right) \Upsilon^1	\nonumber	\\
& &	\qquad +2 v \left(64 i-4\lambda + 3 i \lambda^2 - 2(4+i \lambda ) v - 16 i v^2 \right) {\Upsilon^1}'
	\Bigr]	\; ,	\\
\hspace{-7mm}c^1 &=&
	-\frac{2 v}{\lambda  (4 i - \lambda)}  \Bigl[
	\mathpzc{I}^1
	-(4 i + \lambda - 8 v) \Upsilon^1
	+(4 + i \lambda +8 i v) {\Upsilon^1}'
	\Bigr]	\; .
\eea

\subsection*{{$\Upsilon^2$} :}
\bea
\hspace{-7mm}\Omega^2 &=&
\frac{1}{3} \left[
      \lambda \mathpzc{I}^2
      +(-2+4 i v) \Upsilon^2
      +4 v {\Upsilon^2}'\right]	\; ,	\\
\hspace{-7mm}a^2 &=& \frac{\lambda}{6} \mathpzc{I}^2
      +\frac{-1 + 4 i v}{3} \Upsilon^2
      +v {\Upsilon^2}'	\; ,	\\
\hspace{-7mm} \mathrm{tr}\,h^2 &=&
	-\frac{1}{3} \left[
		\lambda \mathpzc{I}^2
		+(-2+4 i v) \Upsilon^2
		+4 v {\Upsilon^2}'
	    \right]	\; ,	\\
\hspace{-7mm} \hat{h}^2 &=&
	-\frac{1}{3 \lambda} \left[
		\lambda^2 \mathpzc{I}^2
		-\left( \lambda +8 v -8 i v^2 \right) \Upsilon^2
		+2 v (2 i + \lambda +4 v) {\Upsilon^2}'
	     \right]	\; ,	\\
\hspace{-7mm} \hat{\Sigma}^2 &=&
	- \frac{1}{3} \left[
	    \lambda \left(\lambda + 2v \right) \mathpzc{I}^2
	    - 16 v (1-i v) \Upsilon^2
	    +2 v(4 i + \lambda +8 v) {\Upsilon^2}'
	     \right]	\; ,	\\
\hspace{-7mm}\hat{b}^2 &=&
	-\frac{1}{3} \left[
	    \lambda^2  \mathpzc{I}^2
	    -16 v (1-v) \Upsilon^2
	    +2v(4 i + \lambda +8 v) {\Upsilon^2}'
	     \right]	\; ,	\\
& &
\hspace{-7mm}c^2 = \frac{i v}{3} \Upsilon^2 \; .
\eea

\subsection*{{$\Upsilon^3$} :}
\bea
\hspace{-7mm}\Omega^3 &=&
	\frac{1}{12  \left( 4 + \lambda^2 \right)}\Bigl[
	- i \lambda \mathpzc{I}^3	\nonumber	\\
& &	\qquad\qquad + 2\left(i \left( 4 + \lambda^2 \right) - 2\left(4 +8 i \lambda +3 \lambda^2\right) v + 16 \lambda  v^2 \right) \Upsilon^3	\nonumber	\\
& &	\qquad\qquad -4 i v \left(4 + 4i\lambda+\lambda^2 - 8\lambda v \right) {\Upsilon^3}'
	\Bigr]	\; ,	\\
\hspace{-7mm}a^3 &=&
	\frac{1}{ 24 \left( 4 + \lambda^2 \right)}\Bigl[
	-i \lambda \mathpzc{I}^3	\nonumber	\\
& &	\qquad\qquad	2\left(i \left( 4 + \lambda^2 \right) - 8 \left(\lambda^2+2 i \lambda +2\right) v+16 \lambda v^2 \right) \Upsilon^3	\nonumber	\\
& &	\qquad\qquad	- 2v \left( 12 i - 8 \lambda + 3 i \lambda^2 + 16 i v)\right) {\Upsilon^3}'
	\Bigr]	\; ,	\\
\hspace{-7mm} \mathrm{tr}\,h^3 &=&
	\frac{1}{12 \left( 4 + \lambda^2 \right)}\Bigl[
	i \lambda \mathpzc{I}^3	\nonumber	\\
& &	\qquad\qquad	- 2 \left( i \left( 4 + \lambda^2 \right) - 2 \left(4 + 8 i \lambda + 3 \lambda^2 \right) v +16 \lambda v^2 \right) \Upsilon^3	\nonumber	\\
& &	\qquad\qquad	- 2 i v \left( 4 + 4 i \lambda + \lambda^2 - 8 \lambda v \right) {\Upsilon^3}'
	\Bigr]	\; ,	\\
\hspace{-7mm} \hat{h}^3 &=&
	\frac{1}{12 \left( 4 + \lambda^2 \right)}\Bigl[
	i \lambda^2 \mathpzc{I}^3	\nonumber	\\
& &	\qquad\qquad	- \left( i \lambda (4+\lambda^2) + 8 \left( 4i - 3i\lambda^2 - \lambda^3 \right) v - 8 (4-3\lambda^2) \lambda v^2 \right) \Upsilon^3	\nonumber	\\
& &	\qquad\qquad	+2v \left( 8 + 4 i \lambda - 6 \lambda^2 + i \lambda^3 + 4i (4-3\lambda^2) v \right) {\Upsilon^3}'
	\Bigr]	\; ,	\\
\hspace{-7mm} \hat{\Sigma}^3 &=&
	\frac{1}{12 \left( 4 + \lambda^2 \right)}  \Bigl[
	i \lambda \left( \lambda + 2v \right) \mathpzc{I}^3	\nonumber	\\
& &	\qquad\qquad
  + 4v \left( -8 i +2 i \lambda^2 + \lambda^3+8(1+ i \lambda ) v -8 \lambda v^2 \right) \Upsilon^3	\nonumber	\\
& &	\qquad\qquad
  + 2v \left( 16 + 4 i \lambda - 4 \lambda^2 + i \lambda^3 + 8 \left(4i -2\lambda + i \lambda^2\right) v - 32 i \lambda  v^2 \right) {\Upsilon^3}'
	\Bigr]	\; ,	\\
\hspace{-7mm}\hat{b}^3 &=&
	\frac{1}{24 \left( 4 + \lambda^2 \right)}  \Bigl[
    i \lambda^2 \mathpzc{I}^3	\nonumber	\\
& &	\qquad\qquad	- 8v \left( 8 i - 2 i \lambda^2 - \lambda^3 - 4v \left(4-\lambda^2\right) v\right) \Upsilon^3	\nonumber	\\
& &	\qquad\qquad	+ 2v\left( 16 + 4 i \lambda - 4 \lambda^2 + i \lambda^3 + 8 i \left(4-\lambda^2\right) v \right) {\Upsilon^3}'
	\Bigr]	\; ,	\\
\hspace{-7mm} c^3 &=& 
  \frac{v}{12}   \Upsilon^3
 \; .
\eea

\newpage

\section{Green's Functions in Position Space}
\label{app:Fourier}

This appendix deals with some details about the Fourier transformation of the
correlators (\ref{eq:corr_EE_Fourier}-\ref{eq:corr_ElogElog_Fourier})
back to position space.
A useful for this purpose is the Bessel transform,
\bea
\int_0^\infty k^{2\alpha -1} e^{-\beta k^2} J_0(kx) dk
	&=&	\frac{\Gamma(\alpha) e^{-\frac{x^2}{4 \beta}} }{2 \beta^\alpha} {_1F_1}\left(1-\alpha;1;\frac{x^2}{4 \beta}\right)	\; .
\label{eq:power_exp_Bessel_trafo}
\eea
As a direct consequence of this,
\bea
& &	
2^{n} \int_0^\infty k^{2\alpha -1} (\ln k)^n \, e^{-\beta k^2} J_0(kx) \dd k		\nonumber	\\
& & \qquad =	\frac{e^{-\frac{x^2}{4 \beta}}}{2 \beta^\alpha} \sum_{n\geq l \geq 0} \binom{n}{l} (-\ln\beta)^{n-l} \, \partial_\alpha^l \left[ \Gamma(\alpha) {_1F_1}\left(1-\alpha;1;\frac{x^2}{4 \beta}\right) \right]		\; .
\label{eq:power_exp_log_Bessel_trafo}
\eea
Thus, for later convenience, for $n \in \N$ define,
\bea
\mathfrak{l}_n(y)	&=&	\left. \partial_\alpha^n \left[ \frac{\Gamma(\alpha)}{\Gamma(4)} {_1F_1}\left(1-\alpha;1;y\right) \right] \right|_{\alpha=4}	\; .
\label{eq:l_n_def}
\eea
From the properties of confluent hypergeometric functions follows
a recursion relation,
\bea
y {\mathfrak{l}_n}'' + (1-y) {\mathfrak{l}_n}' + 3 {\mathfrak{l}_n} + n \mathfrak{l}_{n-1} &=& 0	\; .
\label{eq:l_n_diffeq}
\eea
This can be used to successively derive expressions for the $\mathfrak{l}_n$ in terms of more
elementary functions,
but they become rather complicated with increasing $n$.
The first two functions are
\bea
\mathfrak{l}_0(y)	&=&	L_3(y)		\; ,	\label{eq:l_0}	\\
\mathfrak{l}_1(y)	&=&	L_3(y) \left[\text{E}_{1}(-y)-\ln (-y)\right]
						- \frac{ \left( y^2 - 8 y + 11 \right)  e^{y} }{6}	
						+\frac{36 y^2  - 180 y + 132}{36}	\; ,
\eea
where $L_3$ denotes the Laguerre polynomial of order $3$ and $\text{E}_{1}$ an exponential integral.
The expression for $\mathfrak{l}_2$ is already a rather lengthy expression involving integrals over
products of exponential and $\text{E}_1$ functions that
will not be written out explicitly here.
It will just be noted that for small values of $y$, the functions $\mathfrak{l}_n(y)$ approach
a finite value that 
can be expressed with polygamma functions and for large values,
\bea
{\mathfrak{l}_n}(y) \xrightarrow{\Re y \to \infty} \frac{6 n (\ln y)^{n-1} e^y}{y^4}	\; .
\label{eq:l_n_asymptotic}
\eea
What is not apparent in this expression, but can be verified by considering that the
definition~\eqref{eq:l_n_def} represents an absolutely converging series, is that the
$\mathfrak{l}_n$ are entire functions on the complex plane and do not contain poles or branch cuts.
Furthermore, by integrating \eqref{eq:l_n_diffeq},
\beq
\int_0^\infty e^{-y} {\mathfrak{l}_n}\, dy
	= -\frac{n}{3} \int_0^\infty e^{-y} {\mathfrak{l}_{n-1}}\, \dd y
	= \ldots
	= \frac{\Gamma(n+1)}{(-3)^n} \int_0^\infty e^{-y} {\mathfrak{l}_{0}}\, \dd y
	= 0	\; ,
\eeq
where last step used \eqref{eq:l_0} and that the $L_n$ form a set of orthogonal polynomials with respect to the weight $e^{-y}$.
Proceeding in a similar fashion,
\beq
\int_0^\infty y^s e^{-y} {\mathfrak{l}_n} \, dy = 0	\quad , \; s = 0,1,2	\;.
\label{eq:l_n_ortho}
\eeq
Now, with regard to (\ref{eq:corr_EE_Fourier}-\ref{eq:corr_ElogElog_Fourier}), define,
\beq
\Psi^{(0)}(\lambda) = \lambda (\lambda^2+4)	\; , \quad
\Psi^{(1)}(\lambda)	= (\lambda^2 + 4) \psi_3(\lambda)  + \lambda \psi_4(\lambda) 	\; , \quad
\Psi^{(2)}(\lambda)	= \psi_3(\lambda)  \psi_4(\lambda) 	\; ,
\eeq
an then consider,
\bea
\mathpzc{h}^{(j)}(t,x)
	&=&	\frac{1}{(2\pi)^3}\int e^{-i\omega t + i k\cdot x} \frac{\omega^2 \, \Psi^{(j)}\left( \frac{k^2}{\omega} \right)}{\Phi\left( \frac{k^2}{\omega} \right)} \dd^2 k \, \dd\omega	\; ,	\nonumber	\\
	&=&	\frac{1}{(2\pi)^2}\int k J_0(kx) e^{-i\omega t} \frac{\omega^2 \, \Psi^{(j)}\left( \frac{k^2}{\omega} \right) }{\Phi\left( \frac{k^2}{\omega} \right)} \dd k \, \dd \omega	\; .
\label{eq:h_j_I}
\eea
where $J_0$ is a Bessel function.
After changing the integration variable from $\omega$ to $\lambda = \frac{k^2}{\omega}$ and
using \eqref{eq:power_exp_Bessel_trafo} to evaluate the integral over $k$,
\bea
\mathpzc{h}^{(j)}(t,x)
	&=&	\frac{1}{(2\pi)^2 t^4}\int k^7 J_0(kx) e^{-\frac{i k^2 t}{\lambda}} \frac{\Psi^{(j)}\left( \lambda \right)}{\lambda^4 \Phi\left( \lambda \right)} \dd k \, \dd \lambda	\; ,	\nonumber	\\
	&=&	\frac{3}{(2\pi)^2 t^4}\int e^{\frac{i \lambda x^2}{4 t}} \mathfrak{l}_0\left(- \frac{i \lambda x^2}{4 t}\right) \frac{\Psi^{(j)}\left( \lambda \right)}{\Phi\left( \lambda \right)} \dd \lambda	\; .
\label{eq:h_j_II}
\eea
It is now manifest that, up to the prefactor, the Green's function depends only on the ratio
$\chi = \frac{x^2}{4 t}$, as it is expected from a theory with
$z=2$ Lifshitz scaling symmetry.
For negative $\chi$, i.e.~$t<0$, the contour can be closed around the lower half-plane
and since
all poles of the integrand have positive imaginary part,
the residue theorem can be used to conclude that it must evaluate to zero.
Thus, 
\bea
\hat{\mathpzc{h}}^{(j)}(t,x)
	&=&	\frac{3 i}{2 \pi} \frac{\theta(t)}{t^4} \mathfrak{h}^{(j)}_0 (\chi)	\, ,	\label{h_final}
\eea
where, in foresight, already for general $s \in \N$,
\beq
\mathfrak{h}^{(j)}_s (\chi)
	=	\frac{1}{2\pi i} \int \limits_{- \bar{\beta}\infty}^{\beta \infty} e^{i \lambda \chi} \mathfrak{l}_s\left( - i \lambda \chi \right) \frac{ \Psi^{(j)}\left( \lambda \right) }{\Phi\left( \lambda \right)}  \, \dd \lambda	\; ,
\label{eq:goh_j_s}
\eeq
with $\Re\beta,\Im\beta > 0$ and the contour is chosen such that all poles of 
the integrand lie above it.
Next, consider the part of the Green's function that contains a logarithm in $\omega$,
\bea
\mathpzc{h}^{(j)}_1(t,x)	&=&	\frac{1}{(2\pi)^2}\int k J_0(kx) \ln(- i \omega) e^{-i\omega t} \frac{\omega^2 \Psi^{(j)}\left( \lambda \right) }{\Phi\left( \lambda \right)} \dd \omega \, \dd k	\; .
\label{eq:h_j_1_I}
\eea
The analysis of this integral can be worked out along the same lines as above, 
i.e.~by first changing the integration variable from $\omega$ to $\lambda$
and then using \eqref{eq:power_exp_log_Bessel_trafo} to integrate over $k$,
\bea
\mathpzc{h}^{(j)}_1(t,x)
&=&	\frac{3}{(2\pi)^2 t^4} \int \left[\ln\left(  \frac{-i}{\lambda} \right) -\ln\left(  \frac{i t }{\lambda} \right)\right]  e^{i \lambda \frac{x^2}{4 t}} \mathfrak{l}_0\left( - i \lambda \frac{x^2}{4 t} \right) \frac{ \Psi^{(j)}\left( \lambda \right) }{\Phi\left( \lambda \right)} \, \dd \lambda 	\nonumber	\\
& &	\qquad + \frac{3}{(2\pi)^2 t^4} \int  e^{i \lambda \frac{x^2}{4 t}} \mathfrak{l}_1 \left( i \lambda \frac{x^2}{4 t} \right) \frac{ \Psi^{(j)}\left( \lambda \right) }{\Phi\left( \lambda \right)} \, \dd \lambda 	\; .
\label{eq:h_j_1_II}
\eea
For $t<0$ the integral over $\lambda$ can again be performed by closing the contour around the lower half-plane and will
therefore evaluate to zero as there are no poles or branch cuts inside the contour.
For $t>0$ there is a branch cut along the positive imaginary axis that needs to be taken into account.
Though, along the real axis $\ln\left(  \frac{-2i}{\lambda} \right) -\ln\left(  \frac{ 2 i }{\lambda} \right) = - \pi i \mathrm{sgn} \lambda$.
Thus, by defining,
\bea
\wt{\mathfrak{h}}^{(j)}_s (\chi)
	&=&	\frac{1}{2\pi i} \int\limits_0^{\beta \infty} e^{i \lambda \chi} \mathfrak{l}_s\left( - i \lambda \chi \right) \frac{ \Psi^{(j)}\left( \lambda \right) }{\Phi\left( \lambda \right)}  \, \dd \lambda
		+\frac{1}{2\pi i} \int\limits_0^{- \bar{\beta} \infty} e^{i \lambda \chi} \frac{ \Psi^{(j)}\left( \lambda \right) }{\Phi\left( \lambda \right)} \mathfrak{l}_s\left( - i \lambda \chi \right) \, \dd \lambda \; ,
\label{eq:tgoh_j_s}
\eea
where the contour is again chosen such that no poles lie below it,
\eqref{eq:h_j_1_I} reduces to
\bea
\mathpzc{h}^{(j)}_1(t,x)
	&=&	\frac{3 i}{2 \pi} \frac{\theta(t)}{t^4} \left[
			-\ln\left( t \right) \, \mathfrak{h}^{(j)}_0(\chi)
			+ \mathfrak{g}_1(\chi) - \pi i  \wt{\mathfrak{h}}^{(j)}_0(\chi)
		\right]	\; .
\label{h_j_1_final}
\eea
It is straightforward to work out that the terms in the Green's functions that 
contain higher powers of logarithms in $\omega$,
\bea
\mathpzc{h}^{(j)}_m (t,x)	&=&	\frac{1}{(2\pi)^2}\int k J_0(kx) \left[\ln(- i \omega)\right]^m e^{-i\omega t} \frac{\omega^2  \Psi^{(j)}\left( \lambda \right) }{\Phi\left( \lambda \right)}\, \dd \omega \, \dd k	\; ,
\label{h_j_m_final}
\eea
translate to polynomial expressions in $\ln t$,
\bea
\mathpzc{h}^{(j)}_m (t,x)
	&=&	\frac{3 i}{2 \pi} \frac{\theta(t)}{t^4} \sum_{m \geq q \geq 0} \binom{m}{q}  \mathfrak{H}_q(\chi) \left( -\ln t \right)^{m-q}	\; .
\eea
The coefficients $\mathfrak{H}^{(j)}_q$ can be expressed in terms of $\mathfrak{h}^{(j)}_s$ and $\wt{\mathfrak{h}}^{(j)}_s$,
\bea
\mathfrak{H}_{2q} 
	&=&	\sum_{q \geq s \geq 0} \binom{2q}{2s}\left(-\pi^2\right)^s \mathfrak{h}^{(j)}_{2(q-s)}
			- \pi i \sum_{q > s \geq 0} \binom{2q}{2s+1}\left(-\pi^2\right)^s \wt{\mathfrak{h}}^{(j)}_{2(q-s)-1}	\; ,	\\
\mathfrak{H}_{2q+1} 
	&=&	\sum_{q \geq s \geq 0} \binom{2q+1}{2s}\left(-\pi^2\right)^s \mathfrak{h}^{(j)}_{2(q-s)+1}
			- \pi i \sum_{q \geq s \geq 0} \binom{2q+1}{2s+1}\left(-\pi^2\right)^s \wt{\mathfrak{h}}^{(j)}_{2(q-s)}	\; .
\eea
An exact analytic evaluation of $\mathfrak{h}^{(j)}_s$ and $\wt{\mathfrak{h}}^{(j)}_s$ for positive values of $t$, respectively $\chi$ could not be obtained.
Nevertheless, it is possible to find estimates for large and small values.
For this purpose, the integrals need to be regularized such that they represent
uniformly convergent expressions.
Therefore, consider,
\bea
\int \limits_{0}^{\infty} e^{i \lambda \chi} \frac{ \Psi^{(j)}\left( \lambda \right) }{\Phi\left( \lambda \right)} \mathfrak{l}_s\left( - i \lambda \chi \right)\, \dd \lambda
&=& \int \limits_{0}^{\beta\infty}  e^{i \lambda \chi} \mathfrak{l}_s\left( - i \lambda \chi \right) \left[ \frac{\Psi^{(j)}\left( \lambda \right) }{\Phi\left( \lambda \right)} - q^{(j)}_{+}(\lambda) \right]\, \dd \lambda \nonumber \\
& &	\qquad + \int\limits_{0}^{\infty} e^{-\varepsilon\lambda}  e^{i \lambda \chi} \mathfrak{l}_s\left( - i \lambda \chi \right) q^{(j)}(\lambda)_{+}\, \dd \lambda \; ,
\label{eq:goh_approx}
\eea
where $\beta$ is again chosen such that no poles lie below the contour and
$q^{(j)}_{+}(\lambda)$ is a function that does neither contain poles
nor branch cuts in the upper half-plane and
$\frac{ \Psi^{(j)}(\lambda) }{\Phi(\lambda)} - q^{(j)}_{+}(\lambda) = o(1/\lambda)$ for
$|\lambda| \gg 1$ in the first quadrant.
Similarly, $q^{(j)}_{-}$ can be defined for the integrals on the negative half axis.
In the case at hand, $q^{(j)}_{\pm}$ can for example be chosen as a combination
of quadratic polynomials and terms involving the $\psi$ function.
The second term in \eqref{eq:goh_approx} can then be analyzed by elementary methods
and it is found that, up to terms involving the $\delta$ distribution,
the contribution from this correction term is at most a constant for $\chi \ll 1$ 
whereas is it is exponentially suppressed for $\chi \gg 1$.

Having established these schemes to regularize the integrands, 
the functions $\mathfrak{h}^{(j)}_s$ and $\wt{\mathfrak{h}}^{(j)}_s$ 
can now be to analyzed
in the aforementioned limits.
For small values of $\chi$, generically,
\bea
\mathfrak{h}^{(j)}_s
	& \stackrel{\chi \to 0}{\longrightarrow} &
		\frac{ \mathfrak{l}_{s}(0) }{\pi i} \, \int
		\limits_{0}^{\infty} \Re \left[ \frac{ \Psi^{(j)}\left( \lambda \right) }{\Phi\left( \lambda \right)} \right]_{reg}\, \dd \lambda 	\; ,	\label{eq:goh_small_chi}	\\
\wt{\mathfrak{h}}^{(j)}_s
	& \stackrel{\chi \to 0}{\longrightarrow} &
		\frac{ \mathfrak{l}_{s}(0) }{\pi} \, \int 
		\limits_{0}^{\infty} \Im\left[ \frac{ \Psi^{(j)}\left( \lambda \right) }{\Phi\left( \lambda \right)} \right]_{reg}\, \dd \lambda	\; .
\eea
What needs to be treated with special care is $\mathfrak{h}^{(0)}_s$.
Since $\left[ \frac{ \Psi^{(0)}\left( \lambda \right) }{\Phi\left( \lambda \right)} \right]_{reg}$
can be chosen such that there are no poles in the lower half plane,
the integral on the right hand side of \eqref{eq:goh_small_chi} would evaluate to zero.
It is however straightforward to expand in $\chi$ and redo the regularization at the first order,
\bea
\mathfrak{h}^{(0)}_s
	& \stackrel{\chi \to 0}{\longrightarrow} &
		- \frac{ \chi \left[ \mathfrak{l}_{s}(0) - \mathfrak{l}'_{s}(0) \right] }{\pi i} \, \int
		\limits_{0}^{\infty} \Im \frac{ \lambda^2 \left( \lambda^2+4 \right) }{\Phi\left( \lambda \right)}\, \dd \lambda 	\; .	\label{eq:goh_0_small_chi}
\eea
For large values of $\chi$, the theorem of residues can be applied to find
\bea
\mathfrak{h}^{(j)}_0
	& \stackrel{\chi \gg 1}{\longrightarrow} &
		\mathrm{Res}_{\lambda_{\min}}\frac{e^{i \lambda \chi} \, L_{3}\left( - i \lambda \chi \right) \Psi^{(j)}\left(\lambda \right) }{\Phi\left( \lambda \right)}
		+ \mathrm{Res}_{-\bar{\lambda}_{\min}}\frac{e^{i \lambda \chi} \, L_{3}\left( - i \lambda \chi \right) \Psi^{(j)}\left(\lambda \right) }{\Phi\left( \lambda \right)}
\; ,
\label{eq:est_large_chi_1}
\eea
where $\{\lambda_{\min},-\bar{\lambda}_{\min}\}$ denotes the pair of zeros
of $\Phi$ with smallest imaginary part.
For $s>0$ the integrand is not any more exponentially suppressed for large values.
Thus, consider,
\bea
\hspace{-10mm}
\int \limits_{0}^{\beta\infty}  e^{i \lambda \chi} \mathfrak{l}_s\left( - i \lambda \chi \right) \left[ \frac{ \Psi^{(j)}\left( \lambda \right) }{\Phi\left( \lambda \right)} \right]_{reg} \, \dd \lambda
&=& \frac{1}{\chi} \int \limits_{0}^{\lambda^{*}\chi}  e^{i \sigma} \mathfrak{l}_s\left( - i \sigma \right)  \left[ \frac{ \Psi^{(j)}\left( \sigma / \chi \right) }{\Phi\left( \sigma / \chi \right)} \right]_{reg} \, d\sigma \nonumber \\
& &	\quad +\frac{1}{\chi} \int \limits_{\lambda^{*}\chi}^{\beta\infty}  e^{i \sigma} \mathfrak{l}_s\left( - i \sigma \right)  \left[ \frac{ \Psi^{(j)}\left( \sigma / \chi \right) }{\Phi\left( \sigma / \chi \right)} \right]_{reg} \, d\sigma \; ,
\label{eq:tgoh_approx}
\eea
where $|\lambda^{*}| < |\lambda_{\min}|$.
Using \eqref{eq:l_n_asymptotic} and \eqref{eq:l_n_ortho} it is then straightforward to conclude that,
\bea
\mathfrak{h}^{(j)}_s
	&\leq & \mathpzc{O} \left( \frac{(\ln \chi)^{s-1}}{\chi^{4}} \right) \; , \quad s \geq 1, \;\chi \gg 1	\; ,
\label{eq:est_large_chi_2}		\\
\wt{\mathfrak{h}}^{(j)}_s
	&\leq & \mathpzc{O} \left( \frac{(\ln \chi)^{\max\{s-1,0\}}}{\chi^{4}} \right) \; , \quad s \geq 0, \;\chi \gg 1	\; .
\label{eq:est_large_chi_tilde_1}
\eea
An exact expression for the proportionality constant on the estimates above could not be obtained with the exception of $s = 0$.
However, as the result looks rather complicated and is not of much relevance in the main section, it will not be mentioned explicitly here.

\newpage

\bigskip
\bibliography{LGbib}

\providecommand{\href}[2]{#2}\begingroup\raggedright\begin{thebibliography}{10}

\bibitem{PhysRevB.14.1165}
J.~A. Hertz, {\it Quantum critical phenomena},  {\em Phys. Rev. B} {\bf 14}
  (Aug, 1976) 1165--1184.

\bibitem{2005Natur.433..226C}
P.~{Coleman} and A.~J. {Schofield}, {\it {Quantum criticality}},  {\em NATURE}
  {\bf 433} (Jan., 2005) 226--229,
  [\href{http://xxx.lanl.gov/abs/cond-mat/0503002}{{\tt cond-mat/0503002}}].

\bibitem{Maldacena:1997re}
J.~M. Maldacena, {\it {The Large N limit of superconformal field theories and
  supergravity}},  {\em Adv.Theor.Math.Phys.} {\bf 2} (1998) 231--252,
  [\href{http://xxx.lanl.gov/abs/hep-th/9711200}{{\tt hep-th/9711200}}].

\bibitem{Hartnoll:2009sz}
S.~A. Hartnoll, {\it {Lectures on holographic methods for condensed matter
  physics}},  {\em Class.Quant.Grav.} {\bf 26} (2009) 224002,
  [\href{http://xxx.lanl.gov/abs/0903.3246}{{\tt arXiv:0903.3246}}].

\bibitem{McGreevy:2009xe}
J.~McGreevy, {\it {Holographic duality with a view toward many-body physics}},
  {\em Adv.High Energy Phys.} {\bf 2010} (2010) 723105,
  [\href{http://xxx.lanl.gov/abs/0909.0518}{{\tt arXiv:0909.0518}}].

\bibitem{Sachdev:2010ch}
S.~Sachdev, {\it {Condensed Matter and AdS/CFT}},  {\em Lect.Notes Phys.} {\bf
  828} (2011) 273--311, [\href{http://xxx.lanl.gov/abs/1002.2947}{{\tt
  arXiv:1002.2947}}].

\bibitem{Kachru:2008yh}
S.~Kachru, X.~Liu, and M.~Mulligan, {\it {Gravity Duals of Lifshitz-like Fixed
  Points}},  {\em Phys.Rev.} {\bf D78} (2008) 106005,
  [\href{http://xxx.lanl.gov/abs/0808.1725}{{\tt arXiv:0808.1725}}].

\bibitem{Balasubramanian:2010uk}
K.~Balasubramanian and K.~Narayan, {\it {Lifshitz spacetimes from AdS null and
  cosmological solutions}},  {\em JHEP} {\bf 1008} (2010) 014,
  [\href{http://xxx.lanl.gov/abs/1005.3291}{{\tt arXiv:1005.3291}}].

\bibitem{Donos:2010tu}
A.~Donos and J.~P. Gauntlett, {\it {Lifshitz Solutions of D=10 and D=11
  supergravity}},  {\em JHEP} {\bf 1012} (2010) 002,
  [\href{http://xxx.lanl.gov/abs/1008.2062}{{\tt arXiv:1008.2062}}].

\bibitem{Cassani:2011sv}
D.~Cassani and A.~F. Faedo, {\it {Constructing Lifshitz solutions from AdS}},
  {\em JHEP} {\bf 1105} (2011) 013,
  [\href{http://xxx.lanl.gov/abs/1102.5344}{{\tt arXiv:1102.5344}}].

\bibitem{Chemissany:2011mb}
W.~Chemissany and J.~Hartong, {\it {From D3-Branes to Lifshitz Space-Times}},
  {\em Class.Quant.Grav.} {\bf 28} (2011) 195011,
  [\href{http://xxx.lanl.gov/abs/1105.0612}{{\tt arXiv:1105.0612}}].

\bibitem{Chemissany:2012du}
W.~Chemissany, D.~Geissbuhler, J.~Hartong, and B.~Rollier, {\it {Holographic
  Renormalization for z=2 Lifshitz Space-Times from AdS}},  {\em
  Class.Quant.Grav.} {\bf 29} (2012) 235017,
  [\href{http://xxx.lanl.gov/abs/1205.5777}{{\tt arXiv:1205.5777}}].

\bibitem{Taylor:2008tg}
M.~Taylor, {\it {Non-relativistic holography}},
  \href{http://xxx.lanl.gov/abs/0812.0530}{{\tt arXiv:0812.0530}}.

\bibitem{Gurarie:1993xq}
V.~Gurarie, {\it {Logarithmic operators in conformal field theory}},  {\em
  Nucl.Phys.} {\bf B410} (1993) 535--549,
  [\href{http://xxx.lanl.gov/abs/hep-th/9303160}{{\tt hep-th/9303160}}].

\bibitem{Flohr:2001zs}
M.~Flohr, {\it {Bits and pieces in logarithmic conformal field theory}},  {\em
  Int.J.Mod.Phys.} {\bf A18} (2003) 4497--4592,
  [\href{http://xxx.lanl.gov/abs/hep-th/0111228}{{\tt hep-th/0111228}}].

\bibitem{Gaberdiel:2001tr}
M.~R. Gaberdiel, {\it {An Algebraic approach to logarithmic conformal field
  theory}},  {\em Int.J.Mod.Phys.} {\bf A18} (2003) 4593--4638,
  [\href{http://xxx.lanl.gov/abs/hep-th/0111260}{{\tt hep-th/0111260}}].

\bibitem{Creutzig:2013hma}
T.~Creutzig and D.~Ridout, {\it {Logarithmic Conformal Field Theory: Beyond an
  Introduction}},  \href{http://xxx.lanl.gov/abs/1303.0847}{{\tt
  arXiv:1303.0847}}.

\bibitem{Bergshoeff:2011xy}
E.~A. Bergshoeff, S.~de~Haan, W.~Merbis, and J.~Rosseel, {\it {A
  Non-relativistic Logarithmic Conformal Field Theory from a Holographic Point
  of View}},  {\em JHEP} {\bf 1109} (2011) 038,
  [\href{http://xxx.lanl.gov/abs/1106.6277}{{\tt arXiv:1106.6277}}].

\bibitem{Ross:2009ar}
S.~F. Ross and O.~Saremi, {\it {Holographic stress tensor for non-relativistic
  theories}},  {\em JHEP} {\bf 0909} (2009) 009,
  [\href{http://xxx.lanl.gov/abs/0907.1846}{{\tt arXiv:0907.1846}}].

\bibitem{Ross:2011gu}
S.~F. Ross, {\it {Holography for asymptotically locally Lifshitz spacetimes}},
  {\em Class.Quant.Grav.} {\bf 28} (2011) 215019,
  [\href{http://xxx.lanl.gov/abs/1107.4451}{{\tt arXiv:1107.4451}}].

\bibitem{FeffermanGraham}
C.~Fefferman and C.~Robin~Graham, {\it {Conformal Invariants}},  in {\em {Elie
  Cartan et les math\'ematiques d'aujourd'hui}}, {Ast\'erisque}, pp.~95--116,
  {Soci\'et\'e Math\'ematique de France, Paris}, June, 1985.
\newblock {hors s\'erie}.

\bibitem{Andrade:2013wsa}
T.~Andrade and S.~F. Ross, {\it {Boundary conditions for metric fluctuations in
  Lifshitz}},  \href{http://xxx.lanl.gov/abs/1305.3539}{{\tt arXiv:1305.3539}}.

\bibitem{Zingg:2011cw}
T.~Zingg, {\it {Thermodynamics of Dyonic Lifshitz Black Holes}},  {\em JHEP}
  {\bf 1109} (2011) 067, [\href{http://xxx.lanl.gov/abs/1107.3117}{{\tt
  arXiv:1107.3117}}].

\bibitem{Baggio:2011cp}
M.~Baggio, J.~de~Boer, and K.~Holsheimer, {\it {Hamilton-Jacobi Renormalization
  for Lifshitz Spacetime}},  {\em JHEP} {\bf 1201} (2012) 058,
  [\href{http://xxx.lanl.gov/abs/1107.5562}{{\tt arXiv:1107.5562}}].

\bibitem{Griffin:2011xs}
T.~Griffin, P.~Horava, and C.~M. Melby-Thompson, {\it {Conformal Lifshitz
  Gravity from Holography}},  {\em JHEP} {\bf 1205} (2012) 010,
  [\href{http://xxx.lanl.gov/abs/1112.5660}{{\tt arXiv:1112.5660}}].

\bibitem{Baggio:2011ha}
M.~Baggio, J.~de~Boer, and K.~Holsheimer, {\it {Anomalous Breaking of
  Anisotropic Scaling Symmetry in the Quantum Lifshitz Model}},  {\em JHEP}
  {\bf 1207} (2012) 099, [\href{http://xxx.lanl.gov/abs/1112.6416}{{\tt
  arXiv:1112.6416}}].

\bibitem{Skenderis:2009nt}
K.~Skenderis, M.~Taylor, and B.~C. van Rees, {\it {Topologically Massive
  Gravity and the AdS/CFT Correspondence}},  {\em JHEP} {\bf 0909} (2009) 045,
  [\href{http://xxx.lanl.gov/abs/0906.4926}{{\tt arXiv:0906.4926}}].

\bibitem{Grumiller:2009mw}
D.~Grumiller and I.~Sachs, {\it {AdS (3) / LCFT (2) ---> Correlators in
  Cosmological Topologically Massive Gravity}},  {\em JHEP} {\bf 1003} (2010)
  012, [\href{http://xxx.lanl.gov/abs/0910.5241}{{\tt arXiv:0910.5241}}].

\bibitem{Grumiller:2009sn}
D.~Grumiller and O.~Hohm, {\it {AdS(3)/LCFT(2): Correlators in New Massive
  Gravity}},  {\em Phys.Lett.} {\bf B686} (2010) 264--267,
  [\href{http://xxx.lanl.gov/abs/0911.4274}{{\tt arXiv:0911.4274}}].

\bibitem{Johansson:2012fs}
N.~Johansson, A.~Naseh, and T.~Zojer, {\it {Holographic two-point functions for
  4d log-gravity}},  {\em JHEP} {\bf 1209} (2012) 114,
  [\href{http://xxx.lanl.gov/abs/1205.5804}{{\tt arXiv:1205.5804}}].

\bibitem{Bergshoeff:2012ev}
E.~A. Bergshoeff, S.~de~Haan, W.~Merbis, J.~Rosseel, and T.~Zojer, {\it {On
  Three-Dimensional Tricritical Gravity}},  {\em Phys.Rev.} {\bf D86} (2012)
  064037, [\href{http://xxx.lanl.gov/abs/1206.3089}{{\tt arXiv:1206.3089}}].

\bibitem{Grumiller:2013at}
D.~Grumiller, W.~Riedler, J.~Rosseel, and T.~Zojer, {\it {Holographic
  applications of logarithmic conformal field theories}},
  \href{http://xxx.lanl.gov/abs/1302.0280}{{\tt arXiv:1302.0280}}.

\bibitem{Son:2002sd}
D.~T. Son and A.~O. Starinets, {\it {Minkowski space correlators in AdS / CFT
  correspondence: Recipe and applications}},  {\em JHEP} {\bf 0209} (2002) 042,
  [\href{http://xxx.lanl.gov/abs/hep-th/0205051}{{\tt hep-th/0205051}}].

\bibitem{Bianchi:2001kw}
M.~Bianchi, D.~Z. Freedman, and K.~Skenderis, {\it {Holographic
  renormalization}},  {\em Nucl.Phys.} {\bf B631} (2002) 159--194,
  [\href{http://xxx.lanl.gov/abs/hep-th/0112119}{{\tt hep-th/0112119}}].

\bibitem{Skenderis:2002wp}
K.~Skenderis, {\it {Lecture notes on holographic renormalization}},  {\em
  Class.Quant.Grav.} {\bf 19} (2002) 5849--5876,
  [\href{http://xxx.lanl.gov/abs/hep-th/0209067}{{\tt hep-th/0209067}}].

\bibitem{Mann:2011hg}
R.~B. Mann and R.~McNees, {\it {Holographic Renormalization for Asymptotically
  Lifshitz Spacetimes}},  {\em JHEP} {\bf 1110} (2011) 129,
  [\href{http://xxx.lanl.gov/abs/1107.5792}{{\tt arXiv:1107.5792}}].

\end{thebibliography}\endgroup
\bibliographystyle{JHEP}

\end{document}